\documentclass[pre,amssymb,twocolumn,showpacs,superscriptaddress,floatfix]{revtex4-1}

\usepackage{graphicx}
\usepackage{bm}
\usepackage{amsmath}
\usepackage{wasysym}
\usepackage[utf8]{inputenc}
\usepackage[T1]{fontenc}
\usepackage{ae,aecompl}

\usepackage{color}
\definecolor{darkblue}{rgb}{0,0,0.6}
\definecolor{darkred}{rgb}{0.6,0,0}

\definecolor{purple}{rgb}{0.4,0,0.3}
\newcommand{\pa}{{\textcolor{purple}{a}}}
\newcommand{\pb}{{\textcolor{purple}{b}}}

\usepackage[colorlinks=true,urlcolor=darkblue,citecolor=darkblue,linkcolor=darkred,hyperfootnotes=false]{hyperref}

\newcommand{\Rbar}{\bar R}
\newcommand{\Cbar}{\bar C}
\newcommand{\Fbar}{\bar F}

\newcommand{\betag}{\ensuremath{\text{\emph\ss}}}
\newcommand{\Dg}{\mathfrak D}
\newcommand{\cg}{\mathfrak c}

\newcommand{\therm}{{\text{th}}}
\newcommand{\lin}{{\text{lin}}}
\newcommand{\resc}{{\text{resc}}}
\newcommand{\toy}{{\text{toy}}}
\newcommand{\Fvar}{\mathcal F_{\text{var}}}
\newcommand{\J}{\mathcal J}

\newcommand{\tho}{{\text{\thorn}} }

\newcommand{\clap}[1]{\hbox to 0pt{\hss#1\hss}}
\newcommand{\mathclapinternal}[2]{\clap{$\mathsurround=0pt#1{#2}$}}
\def\mathclap{\mathpalette\mathclapinternal}

\newcommand{\xitilde}{ \mathclap{\phantom{\;\;}\widetilde{\phantom{i}}} \xi}

\begin{document}

\title{
Finite-temperature and finite-time scaling of the directed polymer free-energy with respect
to its geometrical fluctuations
}

\author{Elisabeth Agoritsas}
\affiliation{DPMC-MaNEP, Universit\'e de Gen\`eve, 24 quai Ernest Ansermet, 1211 Gen\`eve, Switzerland}
\author{Sebastian Bustingorry}
\affiliation{CONICET, Centro At{\'{o}}mico Bariloche, 8400 San Carlos de
Bariloche, R\'{\i}o Negro, Argentina}
\author{Vivien Lecomte}
\affiliation{Laboratoire de Probabilit\'es et Mod\`eles Al\'eatoires (CNRS UMR 7599), Universit\'e Paris VI \& Paris VII, 
site Chevaleret, 175 rue du Chevaleret, 75013 Paris, France }
\author{Gr\'egory Schehr}
\affiliation{Laboratoire de Physique Th\'eorique et Modèles Statistiques (UMR du CNRS 8626), Universit\'e Paris Sud, 91405 Orsay Cedex, France}
\author{Thierry Giamarchi}
\affiliation{DPMC-MaNEP, Universit\'e de Gen\`eve, 24 quai Ernest Ansermet, 1211 Gen\`eve, Switzerland}

\begin{abstract}


%
  We study the fluctuations of the directed polymer in 1+1
  dimensions in a Gaussian random environment with a finite
  correlation length $\xi$ and at finite temperature.
  We address the correspondence between the geometrical transverse fluctuations of
  the directed polymer, described by its roughness, and the
  fluctuations of its free-energy, characterized by its two-point
  correlator.
  Analytical arguments are provided in favor of a generic scaling law
  between those quantities, at finite time, non-vanishing $\xi$ and explicit temperature dependence.
  Numerical results are in good agreement both for simulations on the
  discrete directed polymer and on a continuous directed polymer (with short-range correlated
  disorder).
  Applications to recent experiments on liquid crystals are discussed.

\end{abstract}

\date{\today}

\maketitle
\tableofcontents

\bigskip
\section{Introduction}
\label{sec:intro}


Brownian particles have provided in Physics one of the first example
of systems whose geometrical properties differ radically from those
encountered in regular classical mechanics: instead of following
`smooth' (differentiable) trajectories, those particles follow
continuous but `rough' (non-differentiable) paths due to the
persistent stochastic thermal forces they withstand at finite
temperature.
This phenomenon is described in statistical mechanics by a random
walk, whose geometrical self-similarity at large scale is depicted by
scaling laws and scaling exponents.  These scalings are known to be
distinctive features of universality classes -- gathering different
physical phenomena sharing a common representation.
A natural question regarding such random paths pertains to the
influence of the environment: how does a path in a uniform medium
differ, for instance, from a path in a medium with random
inhomogeneities?  And how do such differences manifest in scaling
properties?

Beyond particle trajectories, those paths also describe generic
interfaces or random manifolds between distinct phases.
Example systems include imbibition
fronts~\cite{moulinet_roughness_2002,alava_imbibition_2004}, wetting
and spreading
interfaces~\cite{bonn_wetting_2009,le_doussal_height_2009}, cracks
propagating in paper~\cite{alava_paper_2006},
avalanches of pinned interfaces~\cite{doussal_statistics_2009,le_doussal_size_2009},
burning
fronts~\cite{maunuksela_kinetic_1997,miettinen_experimental_2005},
interfaces in
magnetic~\cite{lemerle_domain_1998,repain_creep_2004,metaxas_creep_2007}
or ferroelectric
materials~\cite{tybell_domain_2002,paruch_domain_2005} and
generic growth
phenomena~\cite{barabasi_fractal_1995,krug_origins_1997}.
A wide selection of those systems, although ranging from microscopic to macroscopic scale and presenting a large variety of microphysics,
have been described as generic disordered elastic systems
(see Refs.~\cite{blatter_vortices_1994,halpin-healy_kinetic_1995,brazovskii_pinning_2004,
  le_doussal_two-loop_2005,wiese-ledoussal2007,giamarchi_jamming_2006,agoritsas_disordered_ECRYS}
for reviews).

\bigskip

One class of random walk has received close attention in the past decades,
the one-dimensional Kardar-Parisi-Zhang (KPZ)~\cite{KPZ_1986} universality class
(see Refs.~\cite{sasamoto_spohn_2010,corwin_kardarparisizhang_2012} for
recent reviews), since it is related to questions of extremely varied
nature~\cite{kriecherbauer_pedestrians_2010} ranging from Burgers
equation in hydrodynamics~\cite{forster_large-distance_1977}, directed polymers in
random media~\cite{comets_probabilistic_2004,giacomin_random_2007} to
the parabolic Anderson model~\cite{comets_overlaps_2011}, eigenvalues of random
matrices~\cite{johansson_shape_2000,prahofer_universal_2000,herbert_exact_2006},
vicious walkers~\cite{spohn_exact_2006,rambeau_extremal_2010,schehr_extremes_2012,forrester_2011},
dynamics of cold atoms~\cite{kulkarni_gpe_2012} and
transport in 1D stochastic~\cite{gwa_bethe_1992} or deterministic
models~\cite{van_beijeren_exact_2012}.

Here we address the link between the scaling properties of
the \emph{geometry} of the directed polymer (through its
scale-dependent roughness) and the scaling of its \emph{free-energy}
fluctuations.  We examine in particular the rôle of temperature for those scalings and
its possible interplay with finite disorder correlation length,
especially at finite time.

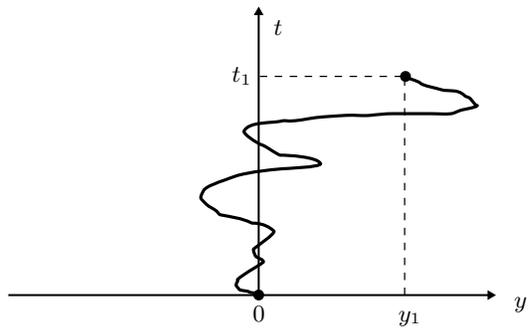
\begin{figure}[!tbp]
\setlength{\unitlength}{.8\columnwidth}
\begin{picture}(1,0.63)%
    \put(0,.06){\includegraphics[width=.75\columnwidth,clip=true]{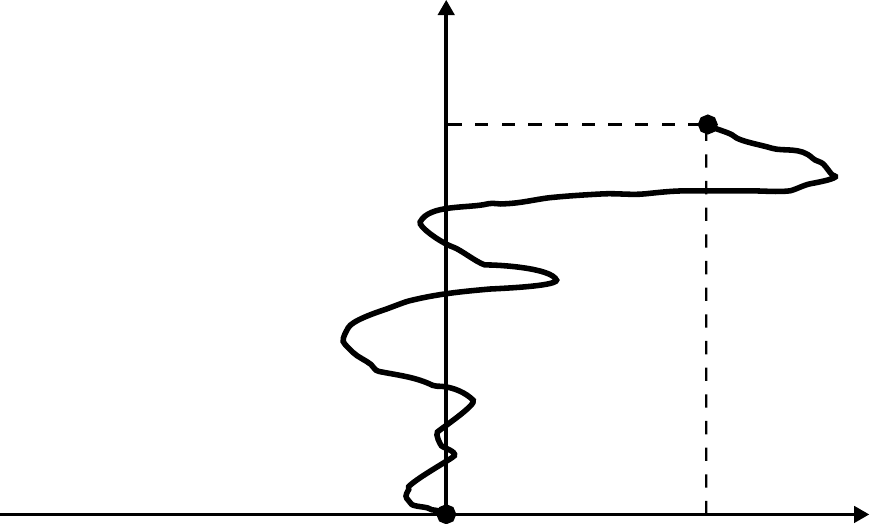}}%
    \put(0.97248501,0.04696){\color[rgb]{0,0,0}\makebox(0,0)[lb]{\smash{$y$}}}%
    \put(0.51052593,0.57){\color[rgb]{0,0,0}\makebox(0,0)[lb]{\smash{$t$}}}%
    \put(0.43,0.48){\color[rgb]{0,0,0}\makebox(0,0)[lb]{\smash{$t_1$}}}%
    \put(0.75,0.02){\color[rgb]{0,0,0}\makebox(0,0)[lb]{\smash{$y_1$}}}%
    \put(0.47,0.02){\color[rgb]{0,0,0}\makebox(0,0)[lb]{\smash{$0$}}}%
  \end{picture}%
\caption{\label{fig:DP_scheme} Schematic view of a continuous directed polymer of trajectory $y(t)$ starting in $y=0$ at time $0$
and arriving in $y_1$ at time $t_1$, in a quenched random potential $V(t,y)$.
These point-to-point configurations correspond to the ``droplet geometry'' of growth models.
}
\end{figure}

More precisely, let us denote by~$t$ the direction of time of the growing
directed polymer (DP), which also represents the `scale' at which a
generic interface is examined, and by~$y$ the transverse direction in
which fluctuations occur (see Fig.~\ref{fig:DP_scheme}).
One important aspect of the KPZ universality class is that
the fluctuations of the (suitably centered) free-energy $\Fbar(t,y)$
are expected to behave as follows in the large time asymptotics:
\begin{equation}
  \label{eq:fluct_Fbar_KPZ}
  \Fbar(t,y) \stackrel{(t\to\infty)}{\sim} a \, t^{\frac 13}\, \chi\Big(\frac{y}{b\,t^{\frac 23}}\Big) \;,
\end{equation}
where $a$ and $b$ are constants which depend on the physical parameters of
the system (\emph{e.g.} temperature~$T$, elasticity~$c$, disorder
strength~$D$ and disorder correlation length~$\xi$ in the model we
use, see section~\ref{sec:model_questions}),
and $\chi(\bar y)$ is a $t$-independent `random variable' whose distribution
characterizes the fluctuations of $y$-dependent observables.
Note that in some systems, \emph{e.g.} liquid crystals, the quantity $\Fbar(t,y)$ scaling
as~\eqref{eq:fluct_Fbar_KPZ} is, up to a translation, the height of an interface and
not its
free-energy~\cite{takeuchi_universal_2010,takeuchi_growing_2011,takeuchi_evidence_2012}.
For a directed polymer with fixed endpoints (as depicted in
Figs.~\ref{fig:DP_scheme} and \ref{f:model}) and with uncorrelated
random environment ($\xi=0$) it has been shown
\cite{prahofer2002,johansson_airy2_2003} that $\chi(y)$, considered as
a (stochastic) function of $y$, is equivalent in distribution to the
so-called Airy$_2$ process (minus a parabola), which is a stationary
determinantal point process. It follows from this
relation~(\ref{eq:fluct_Fbar_KPZ}) that, in the large $t$ limit, the
fluctuations of $\Fbar(t,y)$, at fixed $y$, are described by the
Tracy-Widom distribution ${\cal F}_2$~\cite{tracy_widom_gue_1994}. The
latter distribution ${\cal F}_2$, which can be written as a Fredholm
determinant involving the Airy kernel, describes the fluctuations of
the largest eigenvalue of random matrices belonging to the Gaussian
Unitary Ensemble (GUE). Still at $\xi=0$, remarkably, it was shown
that at finite time $t$, the distribution of $\Fbar(t,y)$, for fixed
$y$, can still be written as a Fredholm determinant, which involves a
non trivial finite $t$ generalization of the Airy
kernel~\cite{calabrese_ledoussal_rosso_2010,amir_corwin_quastel_2011,sasamoto_spohn_exact_2010,dotsenko_bethe_2010}.
%
%

In this paper we focus on the effects of finite temperature and finite disorder
correlation length ($\xi>0$ or discrete DP) at finite time for which we propose a
generalized scaling relation which reads formally
\begin{equation}
  \label{eq:fluct_Fbar_KPZ_sqrtB}
  \Fbar(t,y) \ {\sim}\ \widetilde a \Big[\sqrt{B(t)}\Big]^{\frac 12} 
  \:\chi\Big(\frac{y}{\widetilde b \sqrt{B(t)}}\Big) \;,
\end{equation}
where $B(t)$ is the roughness of the directed polymer, defined as the second cumulant of
the transverse fluctuations along~$y$ at fixed time~$t$.
A more accurate statement is made in
section~\ref{ssec:scaling-arguments} in terms of the two-point correlator of
$\Fbar(t,y)$, see
Eq.~\eqref{eq:scaling_Cyt_with-Bt}.
The large time asymptotics~\eqref{eq:fluct_Fbar_KPZ} is then recovered
from ${B(t) \stackrel{(t\to\infty)}{\sim} t^{\frac 43}}$.
Our motivation for investigating the case~${\xi>0}$ comes from the study
of physical or chemical experiments where the disordered potential
always presents a finite correlation length, which, albeit microscopic
and often inaccessible to direct measurement, may still induce
observable effects at large scales~\cite{agoritsas_disordered_ECRYS}.
In particular, we probe the dependence of the constants $\widetilde a,
\widetilde b$ on the system parameters by distinguishing a high- and a
low-temperature regime due to~$\xi>0$.
The appearance of those regimes generalizes the ones affecting some
specific observables (such as the roughness of the interface, studied
in Refs.~\cite{bustingorry2010,agoritsas2010}).

The directed polymer model we study is defined in section~\ref{sec:model_questions} and
the generalized scaling form we propose is presented and discussed
analytically in section~\ref{sec:scalingCbar}.
Numerical results are gathered in section~\ref{sec:numericalresults_discrete} for the discrete DP
and section~\ref{sec:numericalresults_continuous} for its continuous version.
We discuss experimental implications and draw our conclusions in section~\ref{sec:conclusion}.
The details of some computations are gathered in the appendices.

%


\section{Model and scope of the study}
\label{sec:model_questions}

\subsection{DP formulation}
\label{ssec:model}

\begin{figure}[!tbp]
\setlength{\unitlength}{.75\columnwidth}
\begin{picture}(1,.61)%
    \put(0,.05){\includegraphics[width=\unitlength]{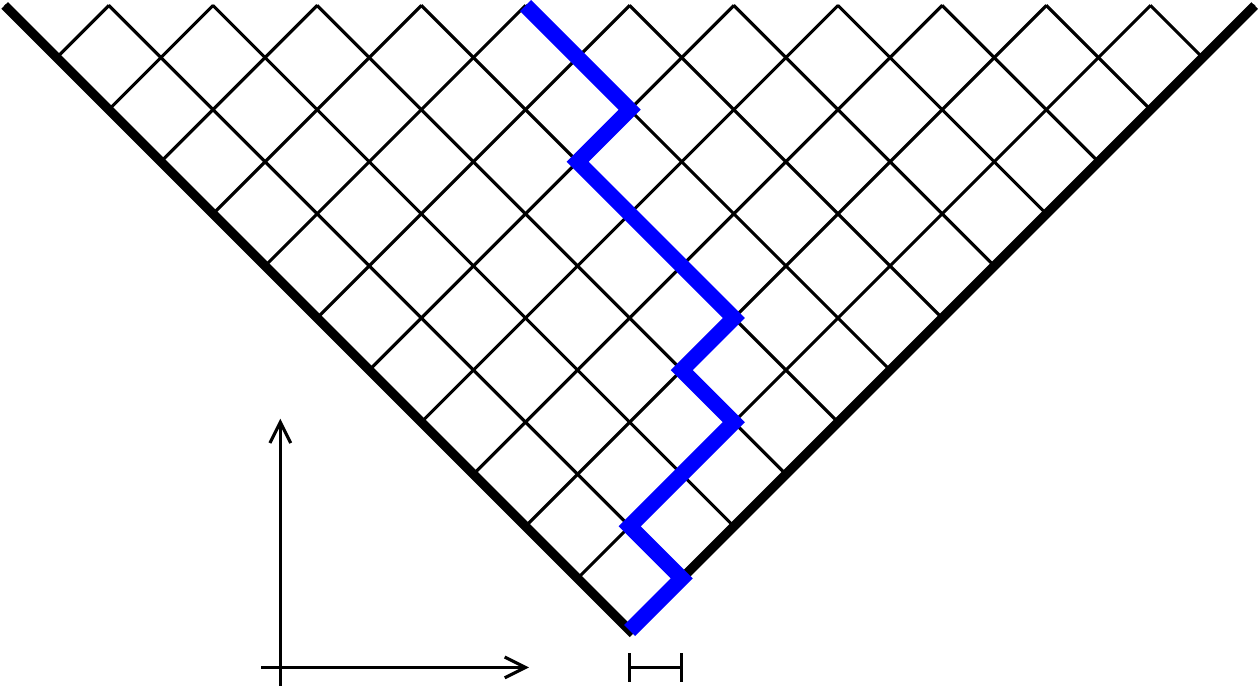}}%
    \put(0.51,0){\makebox(0,0)[lb]{\smash{$1$}}}%
    \put(0.4,0){\makebox(0,0)[lb]{\smash{$y$}}}%
    \put(0.17,.23){\makebox(0,0)[lb]{\smash{$t$}}}%
  \end{picture}%
  \caption{\label{f:model}(Color online) Geometry of the discrete DP model
    we consider. Allowed steps are $y(t+1)-y(t)=\pm 1$. The longitudinal and transverse axes
    correspond to the directions $t$ and $y$ respectively.}
\end{figure}
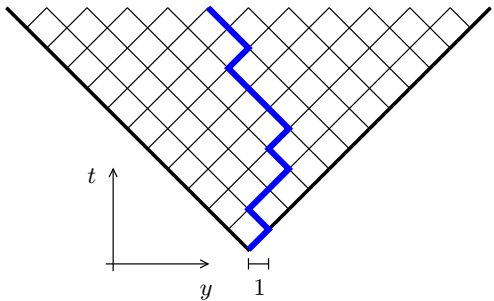

We focus on the directed polymer formulation: the trajectory of the
polymer is described by a continuous coordinate $y(t)$, starting from
$y=0$ at~$t=0$ (see Fig.~\ref{fig:DP_scheme}), and growing in a
random potential $V(t,y)$ along direction~$t$.
The energy of a trajectory $y(t)$ of duration $t_1$ is the sum of elastic and disorder
contributions:
\begin{equation}
 \mathcal H[y,V;t_1] =  \int_0^{t_1} d t\: \Big\{  \frac c2 \big[\partial_ty(t)\big]^2 + V\big(t,y(t)\big)\Big\} \;.
\end{equation}
The first term flattens the polymer by penalizing its deformations
(with an intensity encoded in the elastic constant $c$), while the
second term tends to deform it.
At fixed disorder $V$, the weight of all trajectories starting from
$0$ and arriving in $y_1$ at time $t_1$ is given by the path integral
\begin{equation}
 Z_V(t_1,y_1)= \int_{y(0)=0}^{y(t_1)=y_1} \mathcal Dy(t)\:e^{-\frac 1T \mathcal H[y,V;t_1] } \;,
 \label{eq:defZVty}
\end{equation}
(we set the Boltzmann constant equal to $1$ and denote thereafter the inverse temperature by ${\beta=\frac 1T}$).
The mean value of an observable $\mathcal O$ depending on the value of $y$ 
at final time $t$ reads, 
at fixed disorder
\begin{equation}
  \langle\mathcal O[y(t)]\rangle_V = \frac
 {\int dy\:\mathcal O(y) Z_V(t,y)}
 {\int dy\: Z_V(t,y)} \;.
 \label{eq:averageOZty}
\end{equation}
Here and thereafter, the integrals $\int dy$ over the DP endpoint run
by convention on the real line.
We also consider a discrete version of the same system, illustrated in
Fig.~\ref{f:model} and described in
section~\ref{sec:numericalresults_discrete}.

The weight $Z_V(t,y)$ is not normalized to unity ($\int dy\: Z_V(t,y)\neq 1$ in general)
but the path integral~\eqref{eq:defZVty} is chosen so that at 
zero disorder $\int dy\: Z_{V\equiv 0}(t,y) = 1$ at all times.
With this choice of normalization, it is known from the Feynman-Kac
formula~\cite{feynman_space-time_1948,kac_distributions_1949,HHF_1985,kardar_statistical_2007}
(see also Ref.~\cite{agoritsas_Dtilde_2012} for a discussion) that the
weight $Z_V(t,y)$ evolves according to the `stochastic heat
equation'~\cite{bertini_stochastic_1995,bertini_stochastic_1997,alberts_continuum_2012}
\begin{equation}
  \partial_tZ_V(t,y) = \frac T{2c} \partial_y^2Z_V(t,y) -\frac 1T V(t,y)Z_V(t,y) \;,
  \label{eq:eqevolWpolym}
\end{equation}
while the free-energy
$
  F_V(t,y)=-T \log Z_V(t,y)
$
obeys the Kardar-Parisi-Zhang equation~\cite{HHF_1985,KPZ_1986}
\begin{equation}
  \partial_tF_V(t,y) = \frac T{2c} \partial_y^2F_V(t,y) - \frac 1{2c} \big[\partial_y F_V(t,y)\big]^2+ V(t,y) \;.
  \label{eq:eqevolFpolym}
\end{equation}
The presence of the so-called non-linear `KPZ term' makes the study of
this equation particularly difficult; in particular the typical
extension of the excursions along the direction $y$ does not scale
diffusively at large times (\emph{i.e.} $y \sim t^{\frac 12}$) but
superdiffusively ($y\sim t^{ \zeta}$ where $\zeta=\frac 23$ is the KPZ
exponent)~\cite{forster_large-distance_1977,HHF_1985,KPZ_1986,kardar_replica_1987,johansson_shape_2000,prahofer2002}.

The distribution of the random potential $V(t,y)$ determines the
statistical properties of the free-energy.  We denote by an overline
$\overline{\langle\mathcal O[y(t)]\rangle_V}$ the statistical average
over the disorder~$V$.  One case is well understood: when $V(t,y)$ is a centered
Gaussian uncorrelated random potential~($\xi=0$), that is, when $V(t,y)$ has a
Gaussian distribution with zero mean and two-point correlator
\begin{equation}
  \overline{V(t,y)V(t',y')} = D \delta(t'-t) \delta(y'-y)
  \label{eq:VVcorrelator_deltadelta}
\end{equation}
it is known for long~\cite{HHF_1985} that the infinite-time distribution 
of the free-energy is that of a two-sided Brownian walk:
\begin{equation}
  \text{Prob}[F(y)] \propto
  \exp{\bigg\{-\frac 12 \frac T{cD}\int dy\: \big[\partial_y F(y)\big]^2\bigg\}} \;.
  \label{eq:steadystateKPZ}
\end{equation}
In other words, $F_V(t,y)$ has a steady-state Gaussian distribution 
whose correlator reads:
\begin{equation}
  \overline{\big[F_V(t,y')-F_V(t,y)\big]^2} \xrightarrow[t\to\infty]{ } \frac{cD}T |y'-y| \;.
  \label{eq:Cofyt_infinite-t}
\end{equation}
The value $\frac 23$ for the KPZ exponent is known since the work of
Henley, Huse and Fisher~\cite{HHF_1985} and Kardar, Parisi and
Zhang~\cite{KPZ_1986} and
Kardar~\cite{kardar_roughening_1985,kardar_replica_1987} but this result
has been proven in mathematical
framework much later~\cite{johansson_shape_2000,prahofer2002}.
The free-energy distribution has recently been determined by a variety
of methods both in the physics
~\cite{dotsenko_bethe_2010,calabrese_ledoussal_rosso_2010,dotsenko_klumov_bethe_2010}
and in the
mathematics~\cite{amir_corwin_quastel_2011,sasamoto_spohn_exact_2010,borodin_free_2012}
communities (see Ref.~\cite{corwin_kardar-parisi-zhang_2011} for a
review).

\subsection{Generalization of the free-energy fluctuation scaling}
\label{ssec:motivations}

In the large but finite time limit, Pr\"ahofer and
Spohn~\cite{prahofer2002} have shown that at~$\xi=0$ the correlator
\begin{equation}
 C(t,y)=\overline{\big[F_V(t,y)-F_V(t,0)\big]^2} \;,
 \label{eq:defCty}
\end{equation}
obeys the following scaling relation
\begin{equation}
 C(t,y)\:=\:t^{\frac 23}\,\tilde g_{\text{PS}}\big(yt^{-\frac 23}\big) \;,
 \label{eq:scaling-PS}
\end{equation}
with $\tilde g_{\text{PS}}(y)\sim \frac{cD}{T}|y|$ for small $|y|$ (which
thus gives~\eqref{eq:Cofyt_infinite-t} in the limit $t\to\infty$) while
$\tilde g_{\text{PS}}(y)$ saturates to a constant at large~$|y|$. This function $\tilde g_{\rm PS}(y)$ is, up
to non-universal longitudinal and transverse length scales, the mean square displacement of the Airy$_2$ process
discussed below in Eq.~(\ref{eq:C-Airy2}). 
%
%

Our aim in this paper is to propose and test an extension
of the scaling relation~\eqref{eq:scaling-PS} \emph{(i)} at
finite time and/or \emph{(ii)} for a disorder $V(t,y)$
presenting short-range correlations.
Much less is known in those two cases.
To implement short-range correlations (on a transverse scale~$\xi$) in the
disorder distribution, we assume that it is still zero-mean Gaussian
distributed with correlations of the form
\begin{equation}
  \overline{V(t,y)V(t',y')} = D \delta(t'-t) R_{\xi}(y'-y) \;.
  \label{eq:VVcorrelator_roundeddelta}
\end{equation}
Such correlations in direction $y$ on a range of order $\xi$ are described by
a rounded delta peak $R_\xi(y)$, normalized to unity ($\int dy\,R_\xi(y)=1$)
and by the strength~$D$ of the disorder.
Note that even for $\xi>0$ the distribution of $V$ is invariant by
translation along $y$, as in the uncorrelated
case~\eqref{eq:VVcorrelator_deltadelta} (which corresponds to
$\xi=0$).
For simplicity we may assume in explicit examples that $R_\xi(y)$ is a
normalized Gaussian of variance $2\xi^2$:
\begin{equation}
  R^\text{Gauss}_{\xi}(y) = \frac{1}{\sqrt{4\pi\xi^2}} e^{-\frac{y^2}{4\xi^2}} \;.
  \label{eq:defRtyGauss}
\end{equation}

Note that the knowledge of the fluctuations of the free-energy $F_V$ is not
sufficient to retrieve the mean value of physical observables
from~\eqref{eq:averageOZty}: in general the full distribution of $F_V$
is required.  The free-energy fluctuations described by the correlator $C(t,y)$ still
provide physically relevant information, for instance combined to
scaling arguments~\cite{nattermann_creep}, or as the starting point of `toy-model'
approach where the free-energy distribution is approximated to be
Gaussian~\cite{parisi_replica_1990,mezard1990,bouchaud_bethe_1990,mezard_manifolds_1992,agoritsas2010}
(see also part~\ref{ssec:toymodelwings}).

\section{Scaling of the free-energy correlator $\Cbar_\xi(t,y)$}
\label{sec:scalingCbar}

\begin{figure}[!tbp]
\setlength{\unitlength}{.9\columnwidth}
\begin{picture}(1,0.59574365)%
    \put(0,.06){\includegraphics[width=\unitlength]{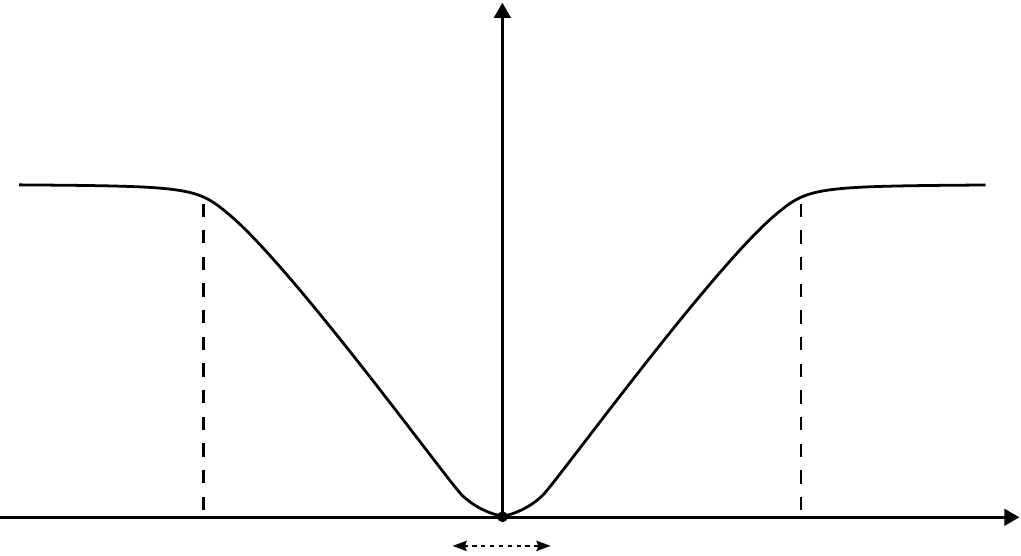}}%
    \put(0.97248501,0.04696){\color[rgb]{0,0,0}\makebox(0,0)[lb]{\smash{$y$}}}%
    \put(0.51052593,0.56407563){\color[rgb]{0,0,0}\makebox(0,0)[lb]{\smash{$\Cbar_\xi(t,y)$}}}%
    \put(0.77530285,0.04696){\color[rgb]{0,0,0}\makebox(0,0)[lb]{\smash{$\ell_t$}}}%
    \put(0.16886934,0.04696){\color[rgb]{0,0,0}\makebox(0,0)[lb]{\smash{$-\ell_t$}}}%
    \put(0.45556981,0.00464578){\color[rgb]{0,0,0}\makebox(0,0)[lb]{\smash{$\sim\xi$}}}%
  \end{picture}%
\caption{Schematic plot of the correlator $\Cbar_\xi(t,y)$ of $\Fbar$,
  as a function of $y$ at fixed $t$. It is rounded close to the origin
  ($|y| \lesssim \xi$), behaves in $\widetilde D |y|$ for intermediate length ($\xi\lesssim |y| \lesssim \ell_t$)
  and goes to a constant at large length ($|y|\gtrsim \ell_t$).
 }
 \label{fig:wingedcorrelator}
\end{figure}
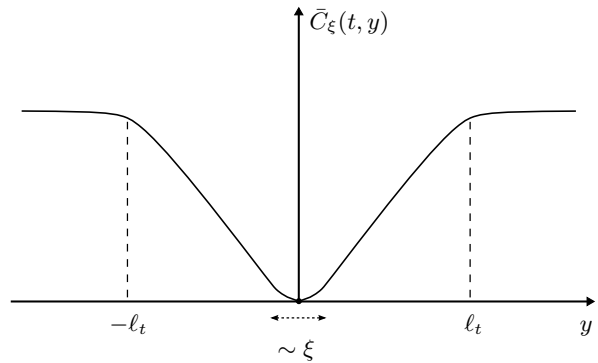

\subsection{A generalized scaling relation}
\label{ssec:generalizedscaling}

Let us first identify a definition of the free-energy correlator which
is suitable to study its finite-time scaling. The correlator
$\overline{[F_V(t,y_2)-F_V(t,y_1)]^2}$ is invariant by translation along
$y$ only in the infinite time limit~\eqref{eq:Cofyt_infinite-t}.
To extend this property at finite time, one may take advantage of the
`statistical tilt symmetry'
(STS)~\cite{fisher_directed_1991,hwa_anomalous_1994,doussal_exact_2003}
of the model
which ensures that the free-energy splits into two contributions (see
\emph{e.g.} Ref.~\cite{agoritsas_Dtilde_2012} for a derivation at non-zero~$\xi$):
\begin{equation}
  F_V(t,y) \ =  \
  \underbrace{c\frac{y^2}{2t} + \frac T2\log \frac{2\pi Tt}c}_{\text{elastic contribution}}
   \ + 
  \underbrace{\,\Fbar_V(t,y)\vphantom{c\frac{y^2}{2t}}\:}_{\substack{\text{disorder}\\ \text{contribution}}}
  \label{eq:defFbar}
\end{equation}
where $c\frac{y^2}{2t} + \frac T2\log \frac{2\pi Tt}c$ is the elastic
contribution, which fully captures the initial condition, while $\bar
F_V(t,y)$ is invariant by translation along $y$ in distribution (in mathematical
terms~\cite{amir_corwin_quastel_2011} $\bar
F_V(t,y)$ is `stationary' along $y$). It represents the contribution
of the disordered potential to the free-energy ($\Fbar_V(t,y)|_{V\equiv 0}=0)$.
Note also from~\eqref{eq:defFbar} that the initial condition
$Z_V(0,y)=\delta(y)$ simply writes $\Fbar_V(0,y)=0$.  This form of
initial condition is technically different from the `sharp wedge'
often considered to pin the polymer in $y=0$ at initial
time~\cite{sasamoto_spohn_exact_2010} and which also ensures
$Z_V(0,y)=\delta(y)$.

This translational invariance allows to define our correlator of interest:
\begin{equation}
 \Cbar_\xi(t,y'-y)= \overline{\big[\Fbar_V(t,y')-\Fbar_V(t,y)\big]^2}
  \label{eq:defCbar}
\end{equation}
where we have made explicit the dependence in $\xi$ arising from the
distribution of~$V$. The STS ensures that~\eqref{eq:defCbar}
depends in $y$ and $y'$ only through the difference $y'-y$,
while the invariance of the distribution of the disorder $V$
by the reflection $V\mapsto -V$ ensures that the function $\Cbar_\xi(t,y)$
is even with respect to its argument $y$.
Note that at $\xi=0$, the steady-state
result~\eqref{eq:Cofyt_infinite-t} implies
\begin{equation}
  \Cbar_{\xi=0}(t,y)\xrightarrow[t\to\infty]{ } \frac{cD}T |y|\:.
  \label{eq:infinitet_zeroxi_Cbar}
\end{equation}
The correlator $ \Cbar_\xi(t,y)$ is actually the connected correlator
of the full free-energy $F_V(t,y)$ (see
appendix~\ref{app:defsCpropsC})
\begin{align}
  \Cbar_\xi(t,y'-y) &= \overline{\big[F_V(t,y')- F_V(t,y)\big]^2} \nonumber \\ &\qquad 
  -\Big[\overline{F_V(t,y')- F_V(t,y)}\Big]^2 \;.
  \label{eq:Cxicorr}
\end{align}
The expression~\eqref{eq:defCbar} is useful to explicit the
translational invariance while~\eqref{eq:Cxicorr} enables to consider
cases where no decomposition such as~\eqref{eq:defFbar} is available
(as for the discrete~DP).

How can we interpret the crossover from the initial condition $\bar
C_\xi(0,y)=0$ to the steady-state result ${\bar C_\xi(\infty,y)\propto
|y|}$?  The weight $\propto e^{-\beta c \frac{y^2}{2t}-\beta
  \Fbar_V(t,y)}$ of a trajectory ending in $y$ at time $t$ depicts a
particle of position $y$ (time $t$ being fixed) in an ``effective
potential'' made up of a parabolic potential describing thermal
fluctuations and of an ``effective disorder'' $\Fbar$,
which recapitulates the disorder landscape $V$ perceived by the
polymer from its starting point.
 Heuristically,
one expects that $\Fbar$ remains almost flat at initial times since
the polymer end-point $y(t)$ has not explored much of its random environment,
while at larger times $\Fbar$ becomes a ``random force potential''
(with cuspy correlator $\Cbar_\xi(t,y)\propto |y|$), at least on a
transverse region $|y|\lesssim \ell_t$ of typical size $\ell_t$ in
which the polymer endpoint has mainly been confined.

We show in appendix~\ref{app:constantClargey} that this intuitive
picture is indeed correct: at all finite times,
${\lim_{|y|\to\infty} \partial_y \Cbar_\xi(t,y)=0}$. The correlator
$\Cbar_\xi(t,y)$ thus has to switch from the absolute value behavior
$|y|$ for $|y|\lesssim\ell_t$ to a plateau for $|y|\gtrsim\ell_t$ (see
Fig.~\ref{fig:wingedcorrelator}), at some scale $\ell_t$ increasing
and diverging with $t$.  We assume that the effect of the correlation
length $\xi$ is to round $\Cbar_\xi(t,y)$ at small $|y|\lesssim\xi$.
The following scaling relation is thus expected to hold:
\begin{align}
  \Cbar_\xi(t,y)\  &= \ \ell_t\,\hat C_{\ell_t^{-1}\!{\xi}}\big(\ell_t^{-1}y\big)
  \label{eq:scalingC_ell-t}
\end{align}
where the scaling function $\hat C_{{\xi}}(y)$ depends on the physical parameters $c$, $D$, $T$ and $\xi$.
At zero $\xi$, it is compatible with the large time
behavior~\eqref{eq:scaling-PS}, provided that $\ell_t\sim t^{\frac 23}$ for
$t\to\infty$.
We now have to identify the crossover length $\ell_t$  at non-zero $\xi$
and finite time.

\subsection{Asymptotic transverse fluctuations}
\label{ssec:scaling-arguments}

%
The variance of the DP endpoint $y(t)$, called the roughness, is the simplest
length quantifying the spatial extension of the polymer
at a given time~$t$: 
\begin{equation}
  B(t) = \overline{\langle y(t)^2\rangle_V} = \overline{
    \ 
    \frac
    {\int dy\: y^2 Z_V(t,y)}
    {\int dy\: Z_V(t,y)}
    \
}\;.
\label{eq:def-roughness_continuousDP}
\end{equation}
It is known that the roughness presents at small times a diffusive
regime ($\zeta_{\text{th}}=\frac12$) and at large times a random
manifold (RM) superdiffusive regime
($\zeta_{\text{RM}}=\frac23$)~\cite{HHF_1985,KPZ_1986,johansson_shape_2000}.
In terms of power laws of the time~$t$, we have the two asymptotic regimes
\begin{equation}
  B(t)\sim \begin{cases}
   t & \text{~for~} t\to 0 \;, \\
   t^{\frac 43} & \text{~for~} t\to \infty \;, \\
  \end{cases} 
\end{equation}
(see also Refs.~\cite{agoritsas2010,agoritsas_Dtilde_2012} when the disorder
correlation length $\xi$ is nonzero).  We propose that $\ell_t\sim\sqrt{B(t)}$ in~\eqref{eq:scalingC_ell-t},
in other words
\begin{align}
  \Cbar_\xi(t,y)\  &= \ \sqrt{B(t)}\, \ \hat C_{\frac{\xi}{\sqrt{B(t)}}}\Big(\tfrac y{\sqrt{B(t)}}\Big) \;.
 \label{eq:scaling_Cyt_with-Bt}
\end{align}
\begin{figure}[!tbp]
\setlength{\unitlength}{.9\columnwidth}
\begin{picture}(1.05,0.7)%
    \put(0,.06){\includegraphics[width=\unitlength]{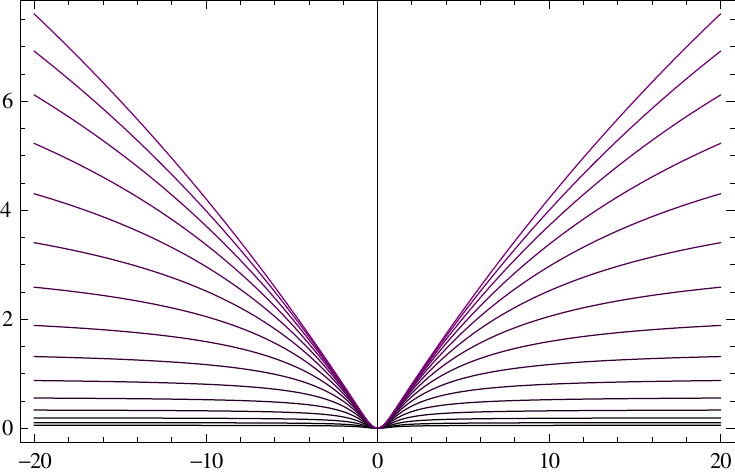}}%
    \put(-0.03,0.35){\color[rgb]{0,0,0}\makebox(0,0)[lb]{\rotatebox{90}{\smash{$\Cbar^{\lin}_\xi(t,y)$}}}}%
    \put(0.5,0.03){\color[rgb]{0,0,0}\makebox(0,0)[lb]{\smash{$y$}}}%
  \end{picture}%
  \caption{(Color online) Graph of the correlator $\Cbar^{\lin}_\xi(t,y)$ of
    $\Fbar$ obtained in the linear
    approximation~\eqref{eq:resCbarlin}, as a function of $y$ for
    different times $t$ (parameters are $c=1$, $D=1$, $T=1$ and $\xi=0.3$).
    Time increases geometrically from bottom ($t=2^{-4}$) to top ($t=2^{10}$) curves.
    }
 \label{fig:wingedcorrelator_linearizeddyn}
\end{figure}

Before testing numerically this scaling law in different models
(see sections~\ref{sec:numericalresults_discrete}
and~\ref{sec:numericalresults_continuous}), we discuss analytical
arguments in favor of~\eqref{eq:scaling_Cyt_with-Bt}.
On one hand, linearizing the dynamics at short time in Eq.~(\ref{eq:eqevolFpolym}) one
finds that the fluctuations are diffusive: $B(t)=\frac{Tt}{c}$. In this
approximation, the correlations of the free-energy rescale as follows
(see Appendix~\ref{sec:shorttime}):
\begin{align}
  \Cbar^\lin_\xi(t,y)\  &= \ \frac {cD}T \sqrt{B_\text{th}(t)}\, \ \hat C^\lin_{\frac{\xi}{\sqrt{B_\text{th\!}(t)}}}\Big(\tfrac y{\sqrt{B_\text{th\!}(t)}}\Big) \;,
\label{eq:scaling_Cyt_with-Bt-linear}
\end{align}
where $B_\text{th}(t)=\frac {Tt}c$ is the thermal roughness. 
The scaling function $\hat C^\lin_{\bar \xi}(\bar y)$, given
in~\eqref{eq:reshatCscalingfunction}, function of the properly
adimensional variables $\bar y$ and $\bar \xi$, is independent of the parameters $c$,
$D$ and $T$.
The behavior of $\Cbar^\lin_\xi(t,y)$ is as qualitatively expected
with a rounding on a scale $\xi$, a plateau at large~$y$ and a
developing linear behavior in between, as plotted in
Fig.~\ref{fig:wingedcorrelator_linearizeddyn}.

On the other hand, at zero $\xi$ and large time, the result~\eqref{eq:scaling-PS} of
Pr\"ahofer and Spohn is compatible
with~\eqref{eq:scaling_Cyt_with-Bt}: this corresponds to the RM
asymptotics $B(t)\sim t^{\frac 43}$.
To address the RM regime at non-zero $\xi$, we study in the next paragraph
a toy-model approach of the DP.

\subsection{A winged DP toy model}
\label{ssec:toymodelwings}

In the infinite time limit and at zero $\xi$, the free-energy $F_V(t,y)$ is
translationally invariant in distribution and has a
Gaussian distribution~\eqref{eq:steadystateKPZ}.  
The statistical tilt symmetry expresses that the free-energy splits at finite time in
two contributions~\eqref{eq:defFbar},
of which only $\Fbar_V(t,y)$ is translationally invariant.
The idea of the
toy
model~\cite{parisi_replica_1990,mezard1990,bouchaud_bethe_1990,doussal_exact_2003}
is to assume that for large~$t$ and finite~$\xi$ the distribution of
the reduced free-energy (denoted $\Fbar^\toy(t,y)$) remains  in
a good approximation Gaussian, with zero mean and correlations
\begin{equation}
 \Cbar_\xi^\toy(t,y'-y)=\overline{\big[\Fbar^\toy(t,y')-\Fbar^\toy(t,y)\big]^2} \;.
 \label{eq:defbarCty}
\end{equation}
It has been shown~\cite{agoritsas2010} that the roughness $B(t)$ can
be computed in a Gaussian Variational Method (GVM) approximation, for
a correlator $\Cbar_\xi^\toy(t,y)$ of the Fourier form
\begin{equation}
  \label{eq:barCtyFourier}
  \Cbar_\xi^\toy(t,y)= \int \frac{d \,\lambda}{2 \pi} \frac 2{\lambda^2} \big[1-\cos (\lambda y)\big] R_{\xitilde}^\text{toy}(t,\lambda) \;,
\end{equation}
with $ R_{\xitilde}^\text{toy}(t,\lambda)= \widetilde D e^{-\lambda^2\xitilde^2 } $.  For
$\xitilde=0$, this writing yields the $\xi=0$ infinite
time-result~\eqref{eq:Cofyt_infinite-t}\, $\Cbar(t,y) = \widetilde D
|y|$, provided that $\widetilde D =\frac{cD}T$.
For $\xi>0$, $R_{\xitilde}^\text{toy}(t,\lambda)$ encodes for $\Fbar$ the finite
correlation length of the disorder $V$: the absolute value becomes rounded
around the origin up to a scale $\sim\xitilde$.
The problem of the form~\eqref{eq:barCtyFourier} of the correlator is
that it implies $\Cbar^\toy_\xi(t,y)\sim \widetilde D |y|$ at large $|y|$, while
we have seen that $ \Cbar_\xi(t,y)$ goes to a constant in the limit
$|y|\to\infty$ at all finite times~$t$.

To overcome the discrepancy between this exact result and the model $
R_{\xitilde}^\text{toy}(t,\lambda)= \widetilde D
e^{-\lambda^2\xitilde^2 } $, one may rather study a correlator
$\Cbar_\xi^\toy(t,y)$ of the form~\eqref{eq:barCtyFourier} with
additional saturation `wings'
\begin{equation}
  R_{\xitilde}^\text{toy}(t,\lambda)= \widetilde D \frac {\lambda^2}{\lambda^2+\ell_t^{-2}}e^{-\lambda^2\xitilde^2 } \;,
  \label{eq:defRlambda_ellt}
\end{equation}
which still presents a behavior $\Cbar^\toy(t,y) \sim \widetilde D |y|$
at intermediate $y$ ($\xi \ll |y| \ll \ell_t$), but goes to a constant
at large $|y|$ ($|y| \gg \ell_t$).
Note that more generically one can consider a toy correlator of the form
\begin{equation}
  R_{\xitilde}^\text{toy}(t,\lambda)= \widetilde D  f_1\big(\lambda\ell_t\big) f_2(\lambda\xitilde ) \;,
  \label{eq:Rtoygeneric}
\end{equation}
in which the rounding due to the finite disorder correlation length appears in the factor $f_2(\lambda\xitilde)$ and 
the crossover at scale $\ell_t$ in the function $f_1\big(\lambda\ell_t\big)$~\footnote{%
   Taking in~\eqref{eq:Rtoygeneric} $f_1(k)\sim k^2$ as $k\to 0$ ensures that $\Cbar^\toy_\xi(t,y)$
   goes to a constant plateau for $|y| \gg \ell_t$ while $f_1(k)\sim 1$ as
   $k\to \infty$ ensures $\Cbar_\xi^\toy(t,y)\sim \tilde D |y| $
   for $\xi\lesssim |y| \lesssim \ell_t$.}.
In particular, scaling properties of the roughness arise from the
form~\eqref{eq:Rtoygeneric}.
Note that this form makes sense  in principle only when the scales $\ell_t$ and $\xitilde$
are separated enough (with $\ell_t>\xitilde$).
We compute in appendix~\ref{app:GVMtoywings} the roughness of this model in the Gaussian Variational
Method (GVM) approximation, in the large $\ell_t$ limit.
The computation yields, in the random manifold regime
\begin{equation}
  B(t) =  B(t)\big|_{\ell_t^{-1}=0} - \frac{\widetilde D}{c^2\ell_t} t^2  + O(\ell_t^{-2}) \;,
\end{equation}
where the RM roughness in the absence of wings ($\ell_t^{-1}$=0) is given by
\begin{equation}
  B(t)\big|_{\ell_t^{-1}=0}  = \frac 32 \Big(\frac{2\widetilde D^2}{\pi c^4}\Big)^{ \frac 13} t^{\frac 43}-\xitilde^2
\end{equation}
For the scale $\ell_t$ not to destroy the $\frac 23$ exponent of the RM regime, one must have
$\ell_t$ growing at least as 
\begin{equation}
  \ell_t \sim \Big(\frac{\widetilde D}{c^2}\Big)^{\frac 13}t^{\frac 23}
\end{equation}
In other words, the wings of $\Cbar(t,y)$ have to appear at a scale
larger than $\sqrt{B(t)}$ in the RM regime.

\section{Numerical simulations at high temperature: the discrete DP}
\label{sec:numericalresults_discrete}

\subsection{Model}

We use numerical simulations of the DP model in order to test the
scaling properties of free-energy fluctuations at finite
temperature. Following the geometry described in Fig.~\ref{f:model} we
perform numerical simulations of a discrete DP model~\cite{kardar_roughening_1985} with the solid-on-solid (SOS)
restriction ${|y(t+1)-y(t)|=1}$. A site-dependent zero-mean
uncorrelated Gaussian disorder potential $V_{t,y}$ of intensity $D$ is used:
\begin{equation}
  \overline{V_{t,y}V_{t',y'}}=D \delta_{t,t'}\delta_{y,y'}
\end{equation}
The energy
of a given configuration of the DP is given by the sum of the site
energies along the path $y(t)$.

Given a disorder realization characterized by $V_{t,y}$, the weight $Z_{t,y}$ of a polymer starting in $(0,0)$ and ending in $(t,y)$ is given by the following recursion:
\begin{equation}
\label{eq:Zeta}
 Z_{t,y} = e^{-\beta V_{t,y}} \left( Z_{t-1,y-1}+ Z_{t-1,y+1}\right)
\end{equation}
with zero time initial condition $Z_{0,y} = \delta_{0,y}$. Therefore,
the probability to observe a polymer ending in $(t,y)$ is
$Z_{t,y}/\sum_{y'} Z_{t,y'}$. Due to the recursion relation, $Z_{t,y}$
grows exponentially with time $t$. To avoid numerical instability, all
weights $Z_{t,y}$ at fixed $t$ are divided by the largest one, which does not
change the polymer ending probability. In terms of the weight
$Z_{t,y}$ the free-energy of the polymer starting in $(0,0)$ and
ending in $(t,y)$ is therefore defined as
\begin{equation}
 F_{t,y} = -T \ln \left( \frac{Z_{t,y}}{\sum_{y'} Z_{t,y'}} \right),
\end{equation}
and the free-energy fluctuations are measured in terms of the connected correlation
\begin{equation}
 \Cbar(t,y) = \overline{\left[ F_{t,y}-F_{t,y'} \right]^2 }-\overline{ \left[ F_{t,y}-F_{t,y'} \right] }^2,
\end{equation}
As discussed in paragraph~\ref{ssec:generalizedscaling}, this
definition is equivalent in the continuum to that involving $\bar
F(t,y)$ in~\eqref{eq:defCbar}.  The index $\xi$ is dropped from
$\Cbar(t,y)$ since the disorder is uncorrelated as
in~\eqref{eq:VVcorrelator_deltadelta}, and since no lengthscale below
the lattice spacing can be considered in this discrete DP model.  The
roughness is defined as the mean square displacement of the free end,
as in~\eqref{eq:def-roughness_continuousDP} for the continuous~DP
\begin{equation}
 \ B(t) = \overline{ \:\langle y(t)^2\rangle\: } =\overline{\sum_y y^2 \frac{Z_{t,y}}{\sum_{y'} Z_{t,y'}}}\:.
\end{equation}

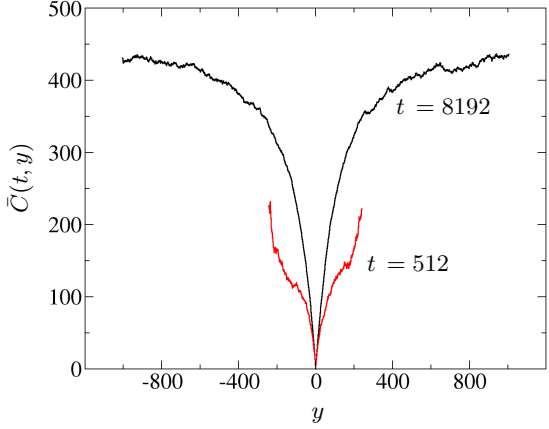
\begin{figure}[!tbp]
\setlength{\unitlength}{.9\columnwidth}
\begin{center}
  \begin{picture}(.9,0.7)
    \put(0,0){\includegraphics[width=.8\columnwidth,clip=true]{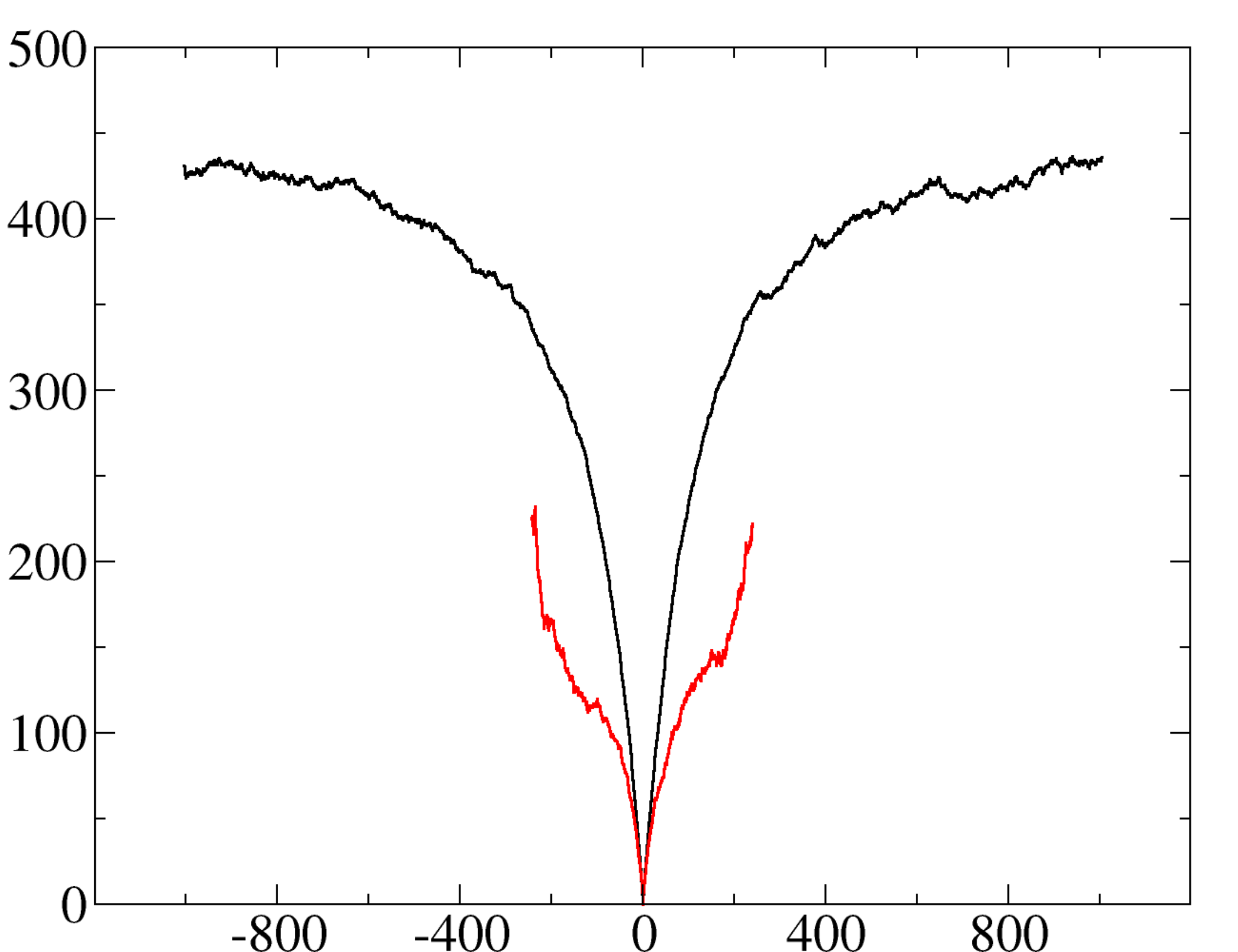}}
    \put(-0.03,0.3){\color[rgb]{0,0,0}\makebox(0,0)[lb]{\rotatebox{90}{\smash{$\Cbar(t,y)$}}}}
    \put(0.455,-0.05){\color[rgb]{0,0,0}\makebox(0,0)[lb]{\smash{$y$}}}
    \put(0.6,0.47){\color[rgb]{0,0,0}\makebox(0,0)[lb]{\smash{$t\,=8192$}}}
    \put(0.55,0.2){\color[rgb]{0,0,0}\makebox(0,0)[lb]{\smash{$t\,=512$}}}
  \end{picture}
\end{center}
\caption{\label{f:CF-finitesize}(Color online) Bare free-energy fluctuations
  $\Cbar(t,y)$ for the discrete DP at finite temperature $T=4$ and for
  two different times $t$. The late increase of $\Cbar(t,y)$ for
  $t=512$ is a finite size effect related to the solid-on-solid
  restriction of the DP model.}
\end{figure}

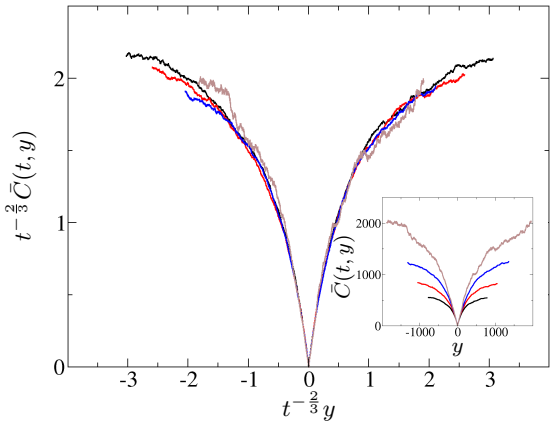
\begin{figure}[!tbp]
\setlength{\unitlength}{.9\columnwidth}
\begin{picture}(.9,0.7)%
    \put(0,0.05){\includegraphics[width=.8\columnwidth,clip=true]{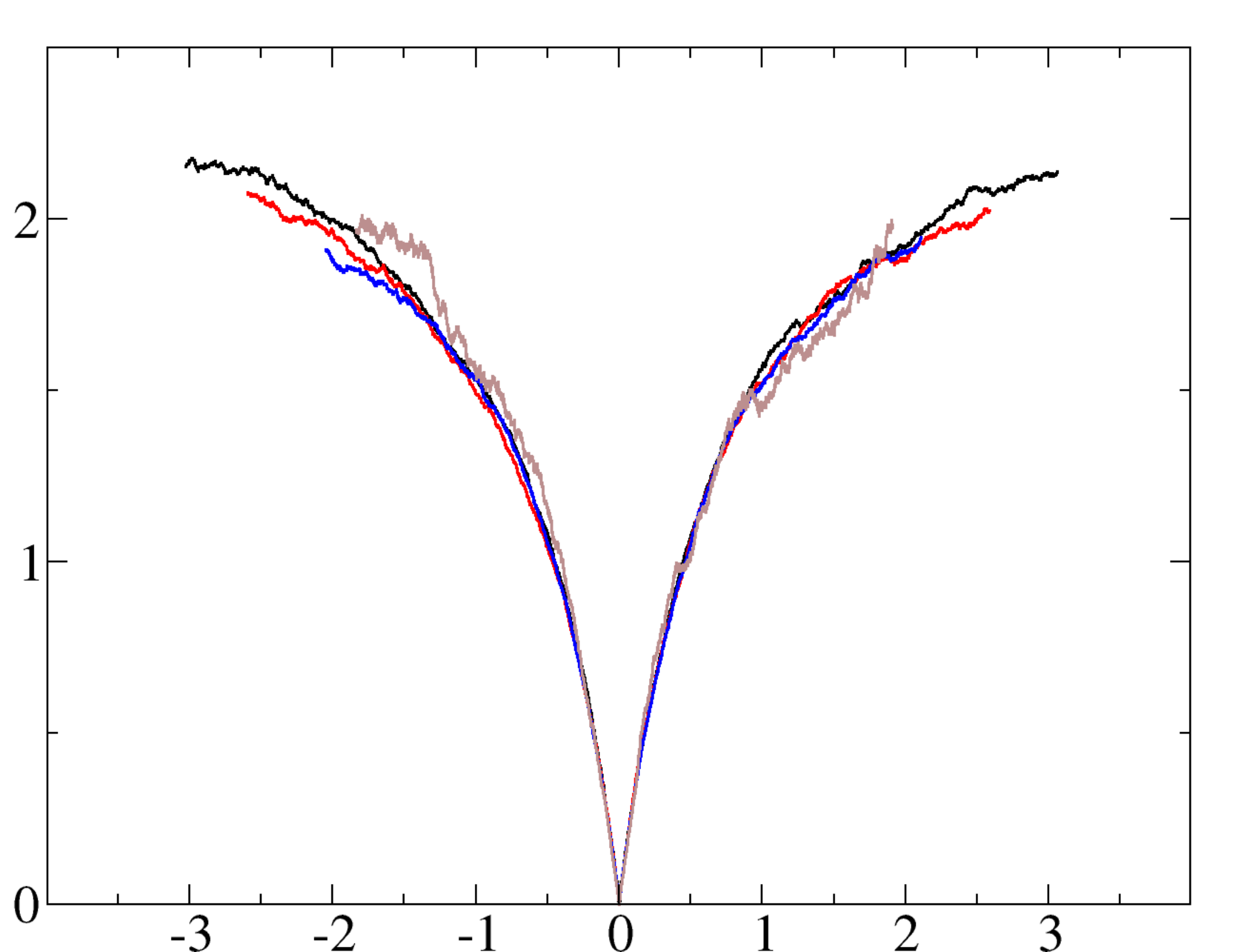}}
    \put(0.52,.13){\includegraphics[width=.3\columnwidth,clip=true]{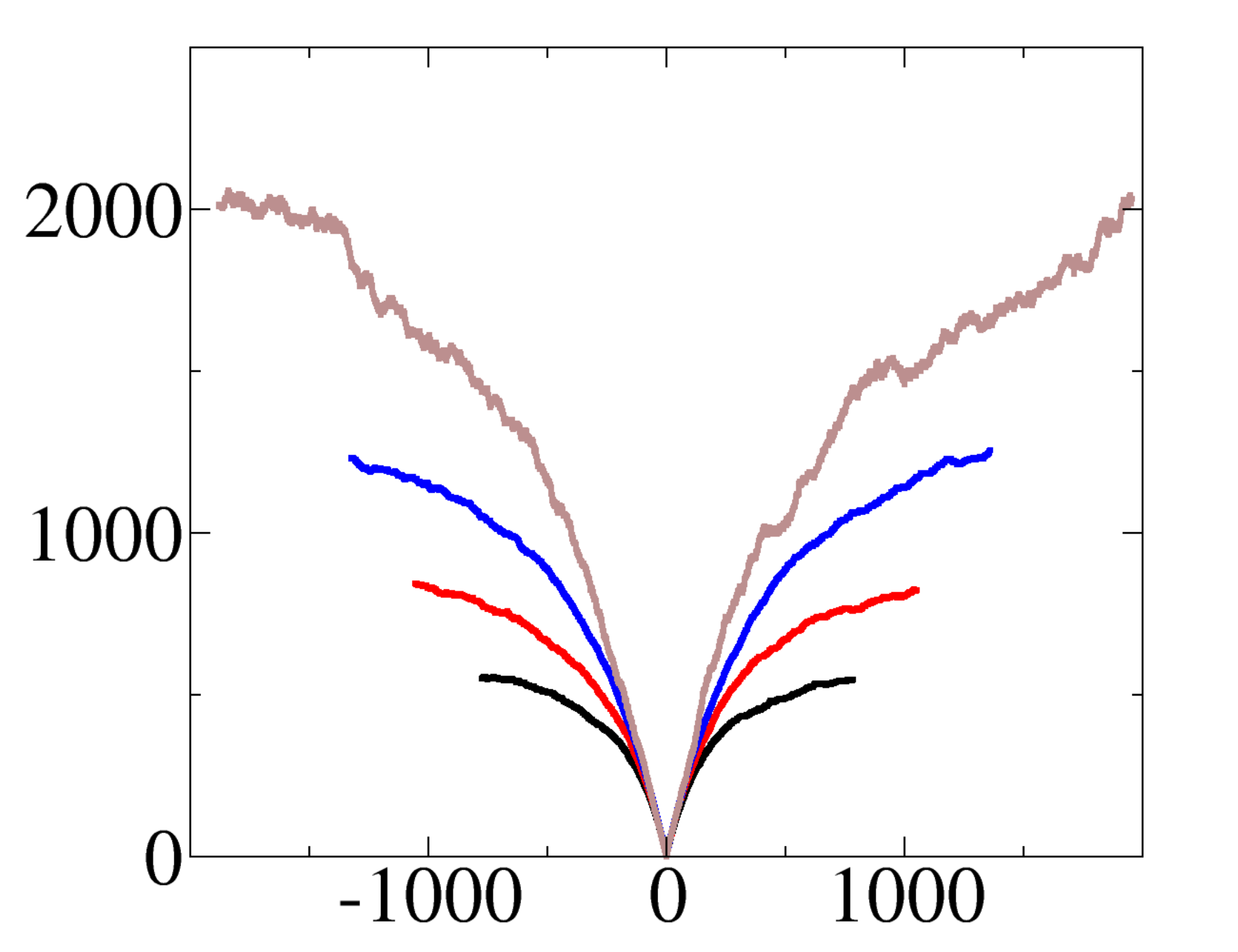}}
    \put(-0.03,0.3){\color[rgb]{0,0,0}\makebox(0,0)[lb]{\rotatebox{90}{\smash{$t^{-\frac 23}\Cbar(t,y)$}}}}%
    \put(0.4,0){\color[rgb]{0,0,0}\makebox(0,0)[lb]{\smash{$t^{-\frac 23}y$}}}%
    \put(0.69,.11){\color[rgb]{0,0,0}\makebox(0,0)[lb]{\smash{\footnotesize$y$}}}%
    \put(0.515,.2){\color[rgb]{0,0,0}\makebox(0,0)[lb]{\rotatebox{90}{\smash{\footnotesize$\Cbar(t,y)$}}}}%
  \end{picture}%
  \caption{\label{f:CF-smallTemp}(Color online) Rescaled free-energy fluctuations
    $\Cbar(t,y)$ for the discrete DP at finite low temperature
    $T=0.5$ and various times, according to the scaling~\eqref{eq:scal-CF-x} with $\zeta=\zeta_\text{RM}=\frac 23$. The inset shows the bare data
    $\Cbar(t,y)$ against $y$ for timescales ${t=4096,8192,16384,32768}$
    from bottom to top.}
\end{figure}

Finally, note the correspondence between the discrete and continuous
parameters [see~\eqref{eq:cbetaDmicmacn1} with lattice spacings $a$ and $b$  taken
to 1)]: the continuum model~\eqref{eq:eqevolWpolym} which
is the limit of the discrete model~\eqref{eq:Zeta} in the large size limit has the parameters
\begin{equation}
 \text{temperature~} T, \quad \text{elasticity~} c=T, \quad \text{disorder~} D
 \label{eq:corresp_cT}
\end{equation}
thus providing a comparison of the temperature dependence of the two
models through the correspondence $c=T$.  A generic study of the
continuum limit of the discrete DP is presented in
Appendix~\ref{app:discretetocontinumDP}.

\subsection{Results}
\label{ssec:result_discrete-DP}

In what follows, results are presented for the discrete DP model with
$D=1$ and disorder averages performed on $10^4$ disorder realizations.
Fig.~\ref{f:CF-finitesize} displays typical curves for the free-energy
fluctuations $\Cbar(t,y)$ at a temperature $T=4$. Two different time
scales are presented, $t=512$ and $t=8192$, in order to show how the
finite size of the system affects free-energy fluctuations.  It is
clear that one observes for both timescales the characteristic $|y|$
behavior at small transverse lengthscales and then a crossover to
saturation at larger timescales. In addition, when the polymer length
is small an increase of $\Cbar(t,y)$ is observed for large $y$, as can
be observed for the $t=512$ data. This last point is a finite-size
effect related to the fact that, in contrast to the continuous case,
for the discrete DP model used here the polymer endpoint is
constrained at all times to remain in the cone ${|y(t)|\leq t}$, see
Fig.~\ref{f:model}. This finite-size effect only plays a role when
analyzing small DP lengths and in the following we discard this
finite-size regime in order to better compare different curves.

We now test the different scaling properties for $\Cbar(t,y)$. 
With respect to time $t$ and in the large time limit~\cite{prahofer2002}, $\Cbar(t,y)$ is expected to scale
according to
\begin{equation}
\label{eq:scal-CF-x}
 \Cbar(t,y) \sim t^{\zeta} \hat C\left( \frac{y}{t^{\zeta}} \right),
\end{equation}
with $\zeta=\zeta_{\text{RM}}=\frac 23$ at large $t$,
which corresponds to the scaling relation in Eq.~\eqref{eq:scaling-PS}.
This behavior has been reported in Ref.~\cite{mezard1990} for a
finite low temperature value ${T\approx 0.14}$ ($\beta = 7$) and moderated polymer sizes $t=512,1024$.
In Fig.~\ref{f:CF-smallTemp} we test the scaling given in
Eq.~\eqref{eq:scal-CF-x} for temperature $T=0.5$, which is larger than
the one used in Ref.~\cite{mezard1990} but still in a low temperature
regime (see below). As shown in the figure the scaling works
satisfactorily for the large time scales used. This result confirms
the scaling probed in Ref.~\cite{mezard1990} and subsequently
analytically obtained in Ref.~\cite{prahofer2002}.

\begin{figure}[!tbp]
\setlength{\unitlength}{.9\columnwidth}
\begin{picture}(.9,0.7)%
    \put(0,0.05){\includegraphics[width=.8\columnwidth,clip=true]{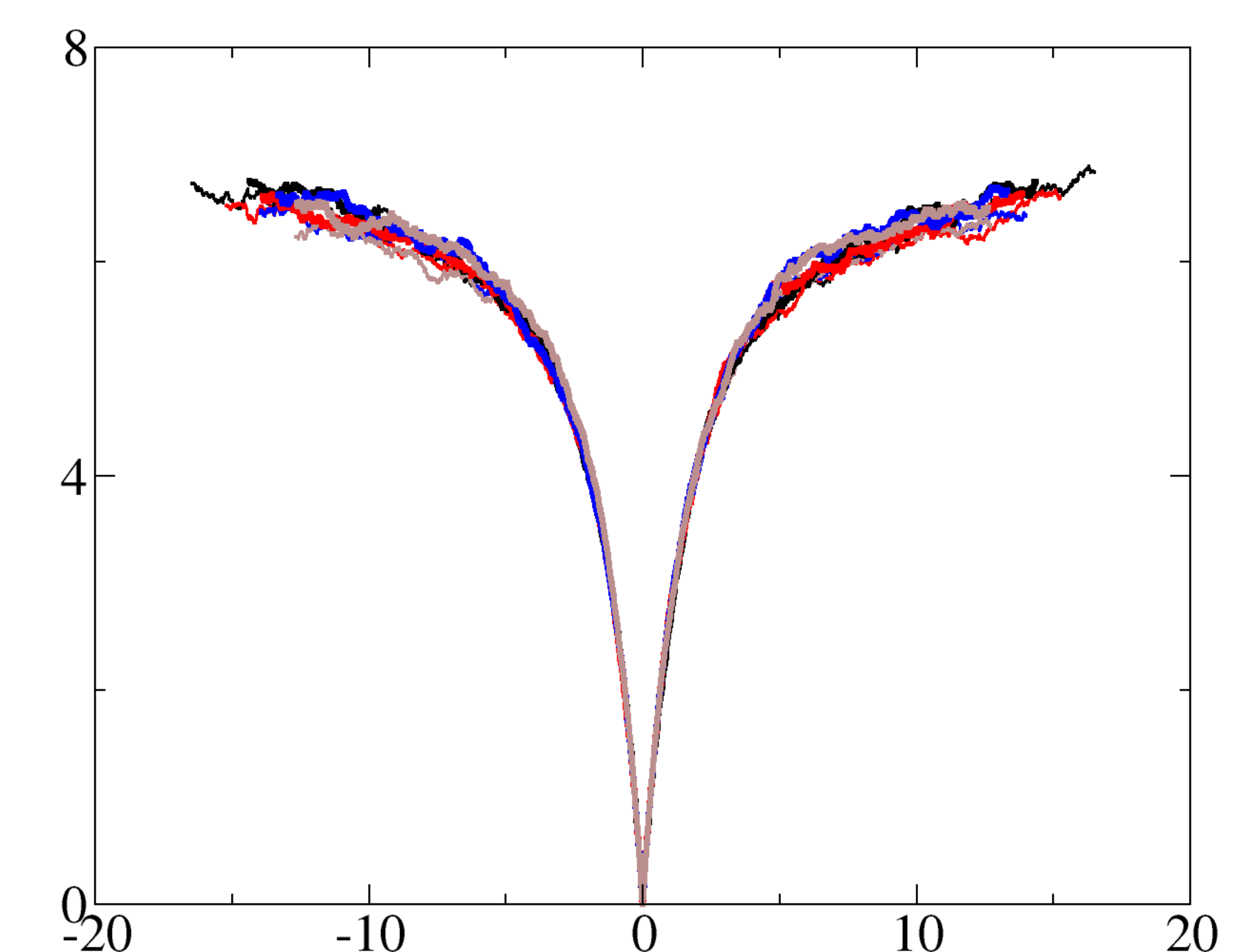}}
    \put(0.52,.13){\includegraphics[width=.3\columnwidth,clip=true]{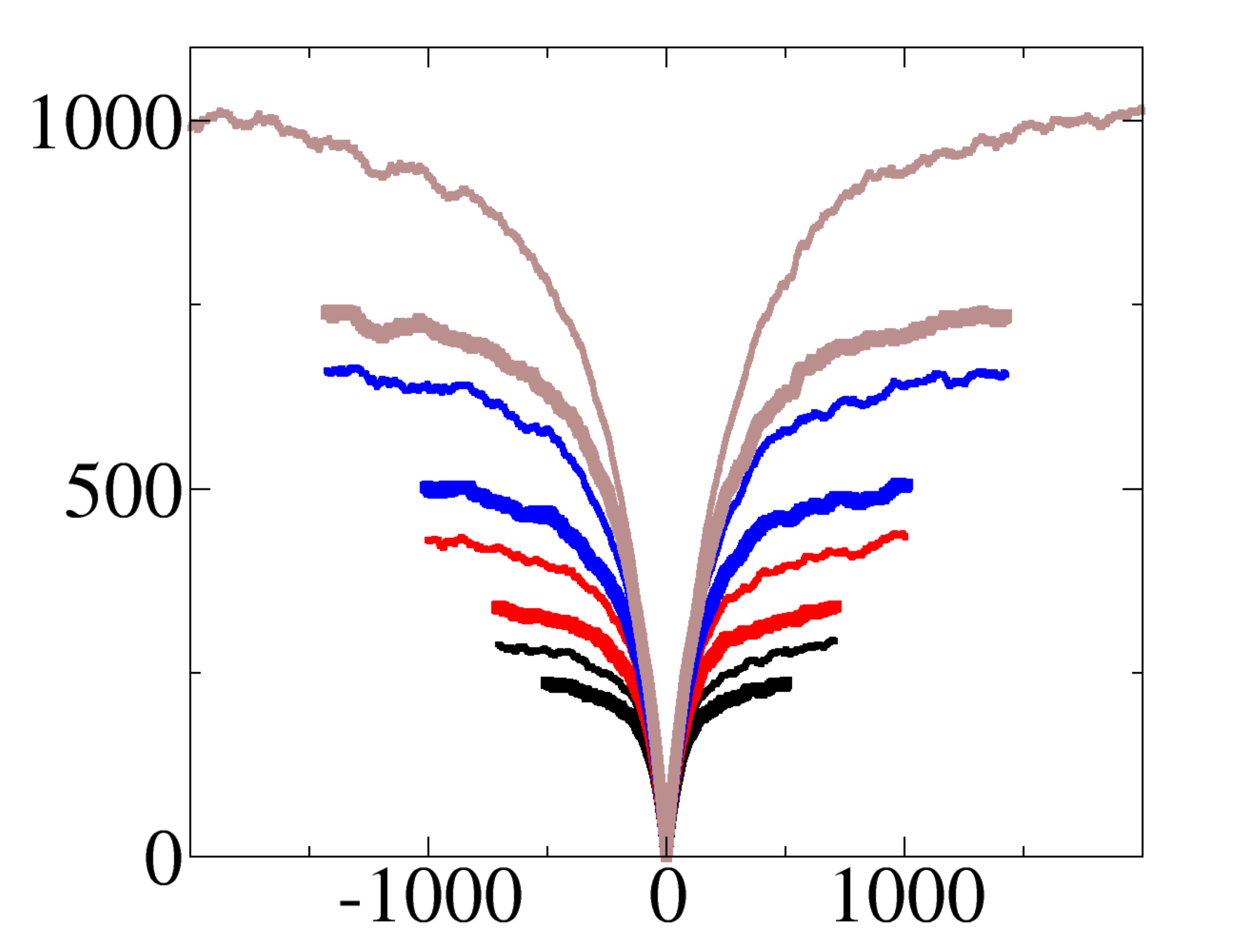}}
    \put(-0.03,0.3){\color[rgb]{0,0,0}\makebox(0,0)[lb]{\rotatebox{90}{\smash{$[B(t)]^{-\frac 12}\Cbar(t,y)$}}}}%
    \put(0.4,0){\color[rgb]{0,0,0}\makebox(0,0)[lb]{\smash{$[B(t)]^{-\frac 12}y$}}}%
    \put(0.69,.11){\color[rgb]{0,0,0}\makebox(0,0)[lb]{\smash{\footnotesize$y$}}}%
    \put(0.52,.2){\color[rgb]{0,0,0}\makebox(0,0)[lb]{\rotatebox{90}{\smash{\footnotesize$\Cbar(t,y)$}}}}%
  \end{picture}%
  \caption{\label{f:CF-T4_8}(Color online) Rescaled free-energy fluctuations
    $\Cbar(t,y)$ for the discrete DP at large times and 
    \textit{high} temperatures, according to the scaling~\eqref{eq:CF-scal-B}. The inset shows the bare data
    $\Cbar(t,y)$ against $y$ for timescales $t=4096,8192,16384,32768$
    from bottom to top. Thin (thick) lines correspond to
    $T=4$ ($T=8$).}
\end{figure}
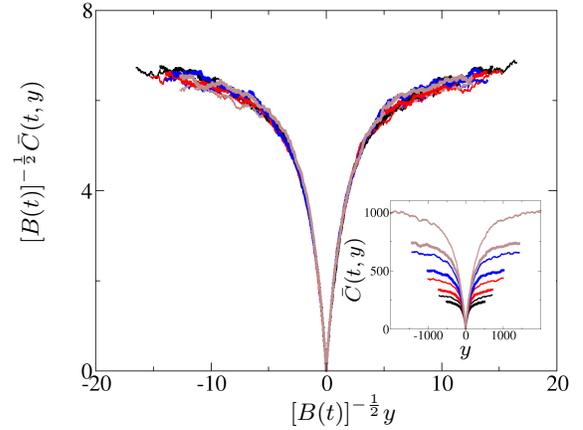

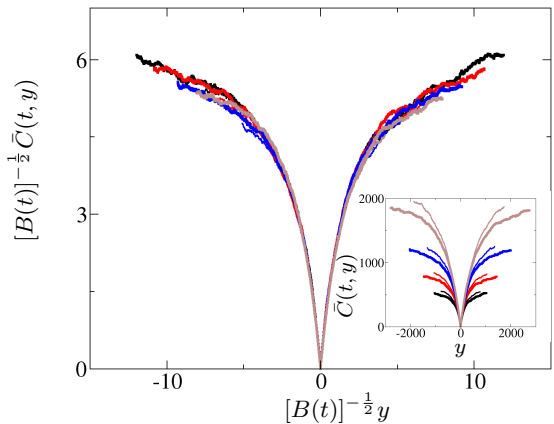
\begin{figure}[!tbp]
\setlength{\unitlength}{.9\columnwidth}
\begin{picture}(.9,0.7)%
    \put(0,0.05){\includegraphics[width=.8\columnwidth,clip=true]{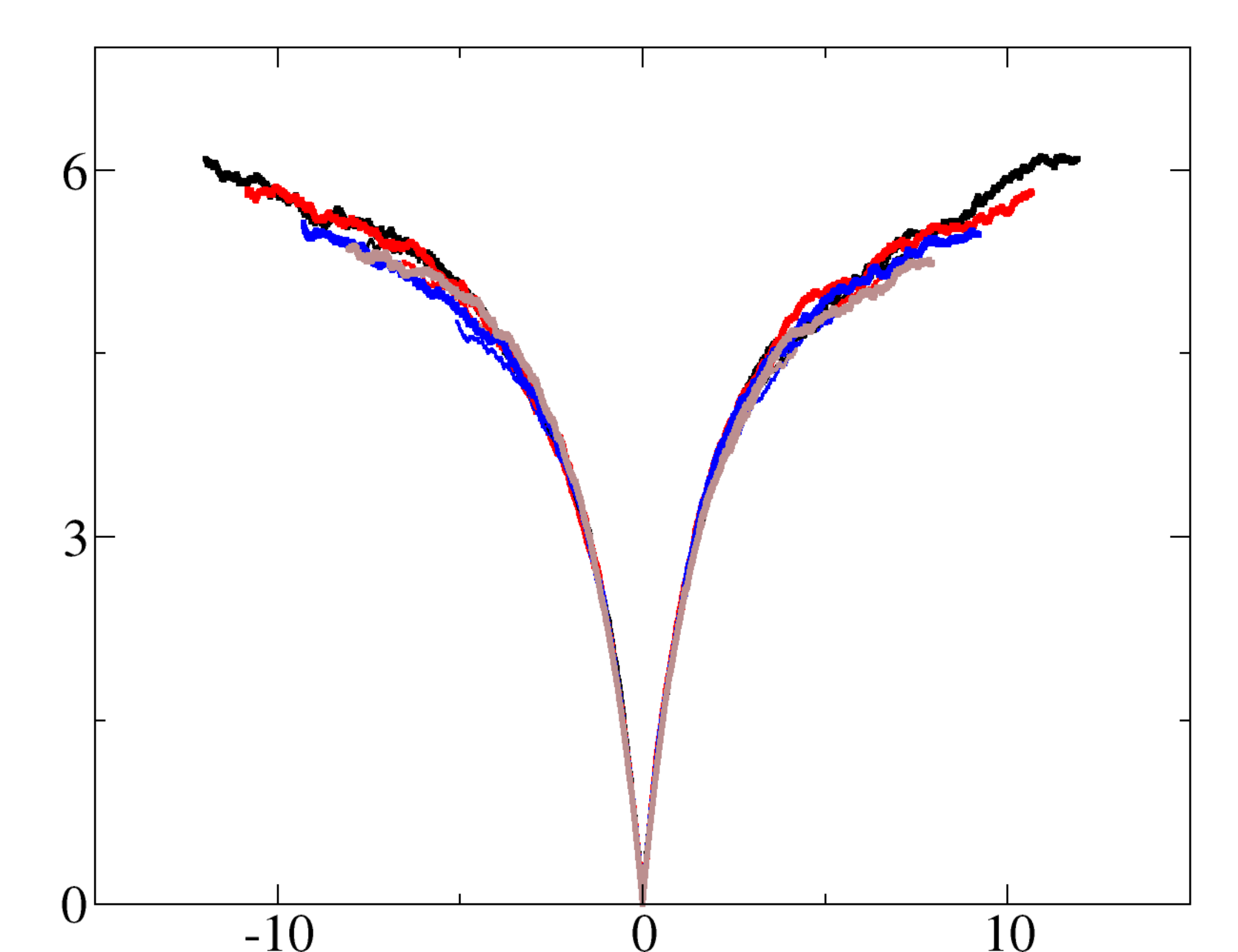}}
    \put(0.521,.13){\includegraphics[width=.3\columnwidth,clip=true]{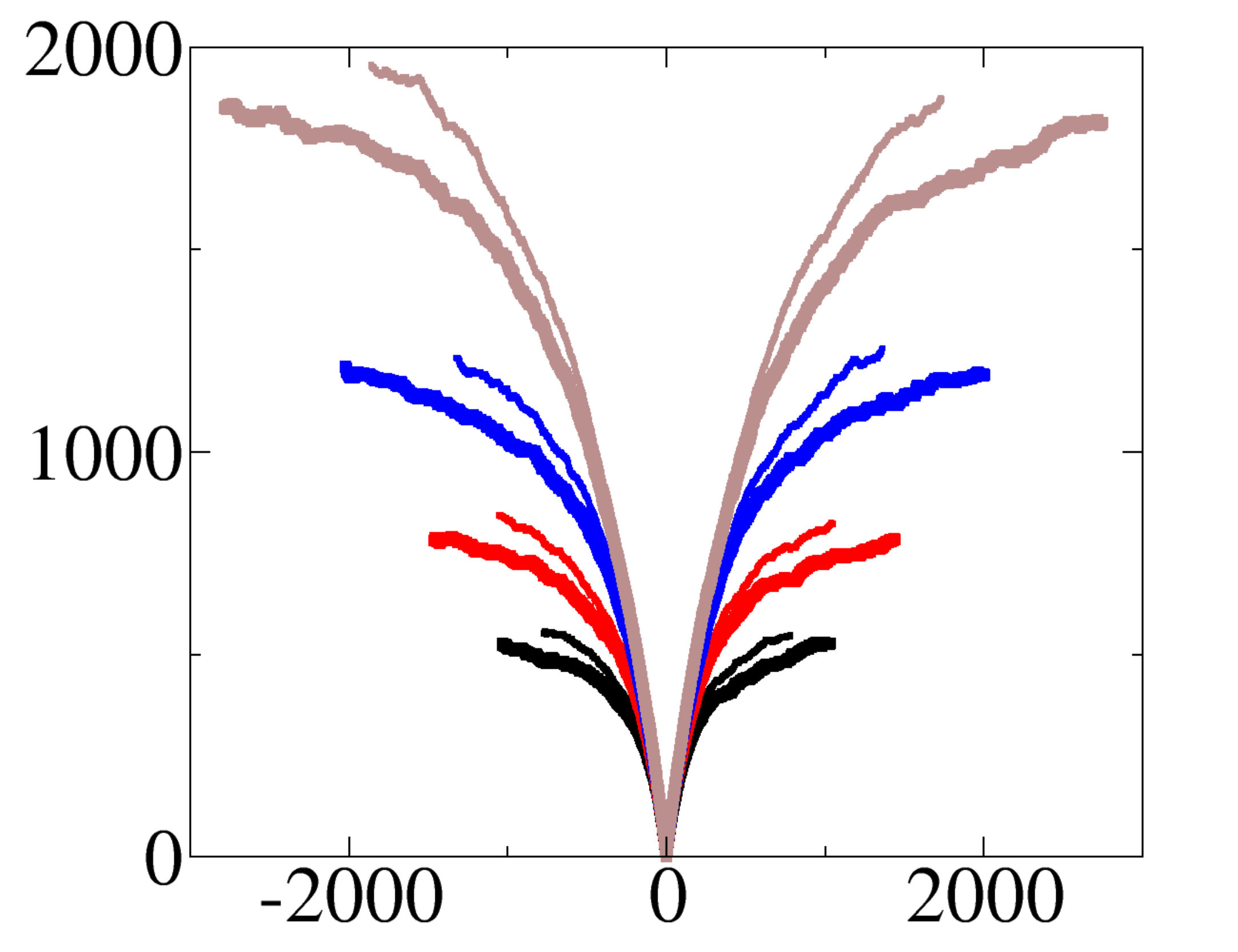}}
    \put(-0.03,0.3){\color[rgb]{0,0,0}\makebox(0,0)[lb]{\rotatebox{90}{\smash{$[B(t)]^{-\frac 12}\Cbar(t,y)$}}}}%
    \put(0.4,0){\color[rgb]{0,0,0}\makebox(0,0)[lb]{\smash{$[B(t)]^{-\frac 12}y$}}}%
    \put(0.691,.11){\color[rgb]{0,0,0}\makebox(0,0)[lb]{\smash{\footnotesize$y$}}}%
    \put(0.516,.17){\color[rgb]{0,0,0}\makebox(0,0)[lb]{\rotatebox{90}{\smash{\footnotesize$\Cbar(t,y)$}}}}%
  \end{picture}%
  \caption{\label{f:CF-T0.5_1}(Color online) Rescaled free-energy fluctuations
    $\Cbar(t,y)$ for the discrete DP at large times and 
    \textit{low} temperatures, according to the scaling~\eqref{eq:CF-scal-B}. The inset shows the bare data
    $\Cbar(t,y)$ against $y$ for timescales $t=4096,8192,16384,32768$
    from bottom to top. Thin (thick) lines
    correspond to $T=0.5$ ($T=1$).}
\end{figure}

In order to go beyond the large time scaling of
Eq.~\eqref{eq:scal-CF-x}, we now discuss the temperature dependence of
free-energy fluctuations with respect to the roughness.  It is first
important to properly consider the finite temperature scaling of the
roughness, which has been discussed in
Refs.~\cite{nattermann_interface_1988,bustingorry2010,agoritsas2010}. In
the high temperature regime and within the continuum limit of the DP, the
roughness scales asymptotically as
\begin{equation}
 B(t) \sim
  \label{eq:Bscal-highT_with-c}
  \left\{
   \begin{array}{ll}
    \frac{T}{c} t^{2\zeta_{\text{th}}} &  (t \ll L_T)\\
     \big(\frac{c T}{D}\big)^{2\tho} t^{2\zeta_\text{RM}} & (t \gg L_T)
   \end{array}
  \right.
  \text{with~} L_T=\frac{T^5}{cD^2},
\end{equation}
where $\zeta_{\text{th}}=1/2$ is the thermal roughness exponent and
$\zeta_\text{RM}=\frac 23$. The temperature dependence of the
roughness is described at large scale by the \textit{thorn} exponent
${\tho=-1/3}$. The thermal lengthscale $L_T$ separates short-scale
thermal fluctuations characterized by~$\zeta_{\text{th}}$ from
large-scale disorder-induced fluctuations of
exponent~$\zeta_\text{RM}$. It has been shown~\cite{nattermann_interface_1988} that in the
high temperature regime the thermal length scale grows as $L_T \sim
T^{1/\theta_F}/(cD^2)$, where $\theta_F=1/5$ is the Flory
exponent~\cite{bustingorry2010,agoritsas2010}. In the discrete version
of the DP one has $c=T$ [see~\eqref{eq:corresp_cT} and
Appendix~\ref{app:discretetocontinumDP}] and therefore the scaling of
the roughness~\eqref{eq:Bscal-highT_with-c} now reads
\begin{equation}
 B(t) \sim
  \label{eq:Bscal-highT}
  \left\{
   \begin{array}{ll}
    t^{2\zeta_{\text{th}}} &  (t \ll L_T)\\
     \big(\frac{T^2}{D}\big)^{2\tho} t^{2\zeta_\text{RM}} & (t \gg L_T)
   \end{array}
  \right.
  \text{with~} L_T=\frac{T^4}{D^2}\;.
\end{equation}
In this last case, remarkably, the short-time behavior is temperature-independent
while the large time prefactor scales with temperature
as $T^{4\tho}= T^{-4/3}$.

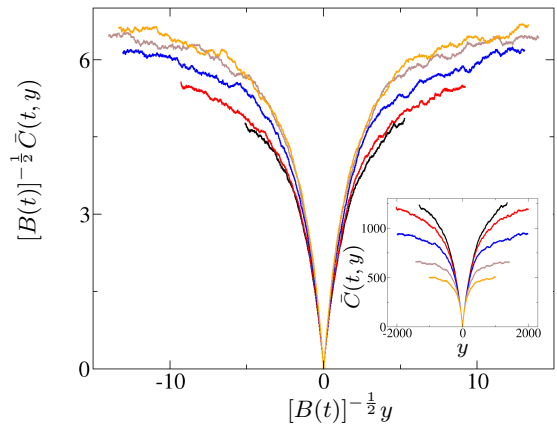
\begin{figure}[!tbp]
\setlength{\unitlength}{.9\columnwidth}
\begin{picture}(.9,0.7)%
    \put(0,0.05){\includegraphics[width=.8\columnwidth,clip=true]{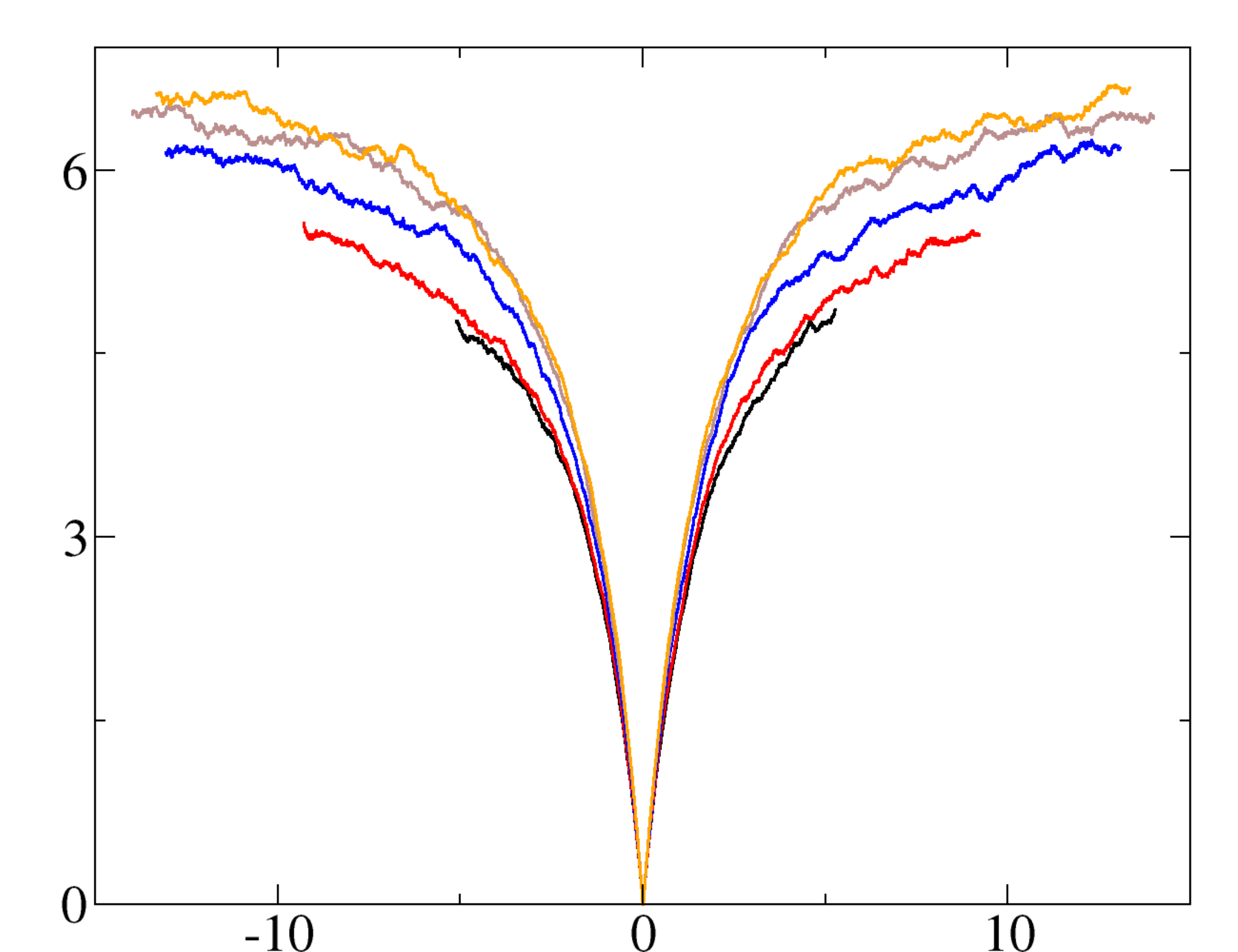}}
    \put(0.52,.13){\includegraphics[width=.3\columnwidth,clip=true]{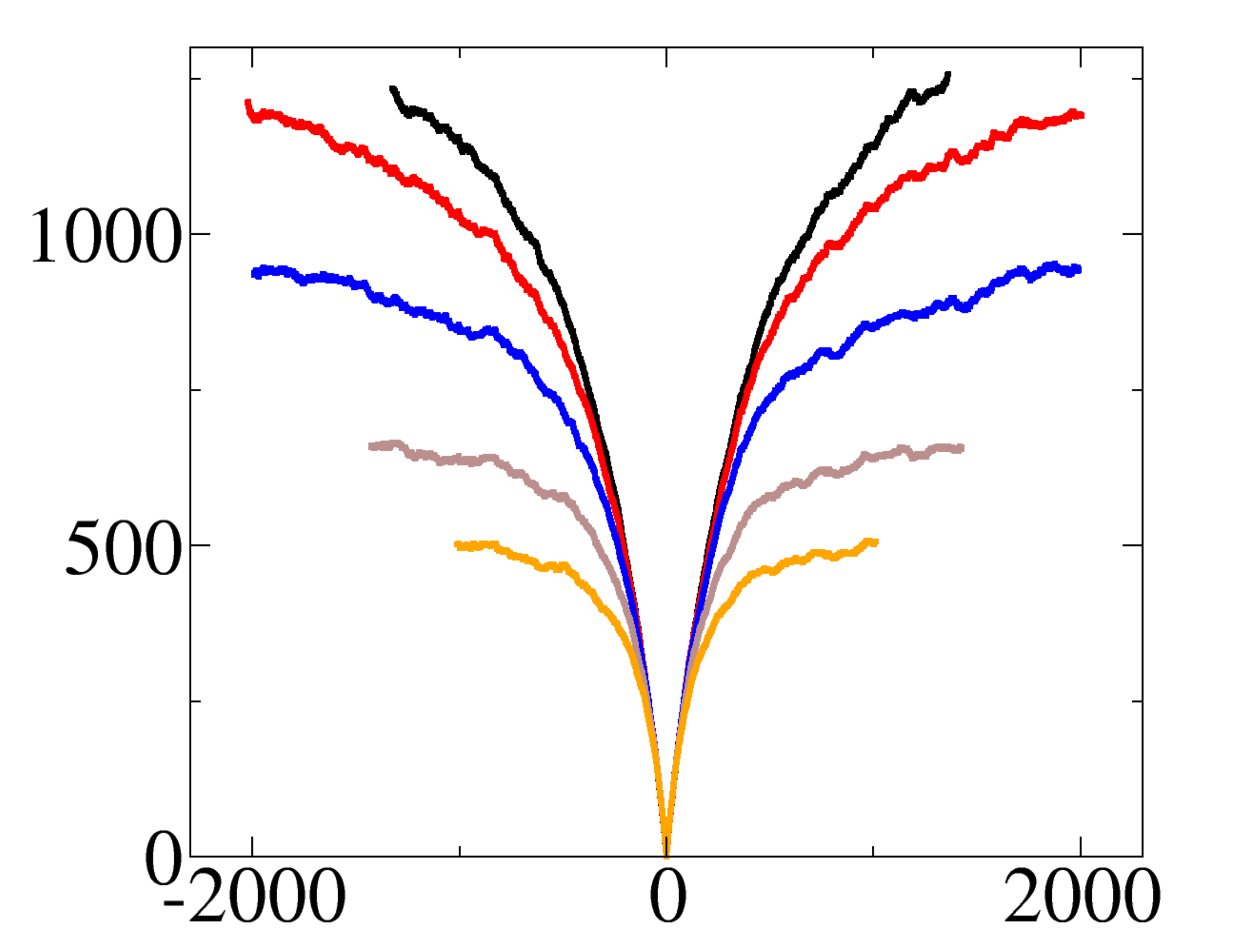}}
    \put(-0.03,0.3){\color[rgb]{0,0,0}\makebox(0,0)[lb]{\rotatebox{90}{\smash{$[B(t)]^{-\frac 12}\Cbar(t,y)$}}}}%
    \put(0.4,0){\color[rgb]{0,0,0}\makebox(0,0)[lb]{\smash{$[B(t)]^{-\frac 12}y$}}}%
    \put(0.69,.11){\color[rgb]{0,0,0}\makebox(0,0)[lb]{\smash{\footnotesize$y$}}}%
    \put(0.523,.185){\color[rgb]{0,0,0}\makebox(0,0)[lb]{\rotatebox{90}{\smash{\footnotesize$\Cbar(t,y)$}}}}%
  \end{picture}%
  \caption{\label{f:CF-t16384}(Color online) Rescaled free-energy fluctuations
    $\Cbar(t,y)$ for the discrete DP at a fixed large time $t=16384$
    and from low to high temperatures, according to the
    scaling~\eqref{eq:CF-scal-B}. The inset shows the bare data
    $\Cbar(t,y)$ against $y$. Temperatures are
    $T=0.5,1,2,4,8$ from top to bottom.}
\end{figure}

These scaling properties of the roughness can be directly incorporated
into the scaling properties of free-energy correlations using the whole
roughness to define the characteristic transverse scale as in the scaling
relation~\eqref{eq:scaling_Cyt_with-Bt} which, in the absence of
$\xi$, writes
\begin{equation}
 \label{eq:CF-scal-B}
 \Cbar(t,y) = \sqrt{B(t)} \; \hat C\Big( \tfrac{y}{\sqrt{B(t)}} \Big).
\end{equation}
First note that the inset of Fig.~\ref{f:CF-t16384} shows how the
saturation regime of free-energy fluctuations is reached at smaller
transverse lengthscale $y$ when increasing the temperature. If, as
suggested by the scaling relation~\eqref{eq:CF-scal-B}, the crossover
is dictated by the time dependent roughness $B(t)$, this is in
agreement with the negative value of the \textit{thorn} exponent
ruling the temperature dependence of the
roughness~\cite{nattermann_interface_1988,bustingorry2010,agoritsas2010}. The
main panel of Fig.~\ref{f:CF-T4_8} illustrates the validity of the
scaling relation~\eqref{eq:CF-scal-B} that incorporates the
temperature dependence through the roughness function, in the range of
high-temperature roughness.

\begin{figure}[!tbp]
\setlength{\unitlength}{.9\columnwidth}
\begin{picture}(.9,0.7)%
    \put(0,0.05){\includegraphics[width=.8\columnwidth,clip=true]{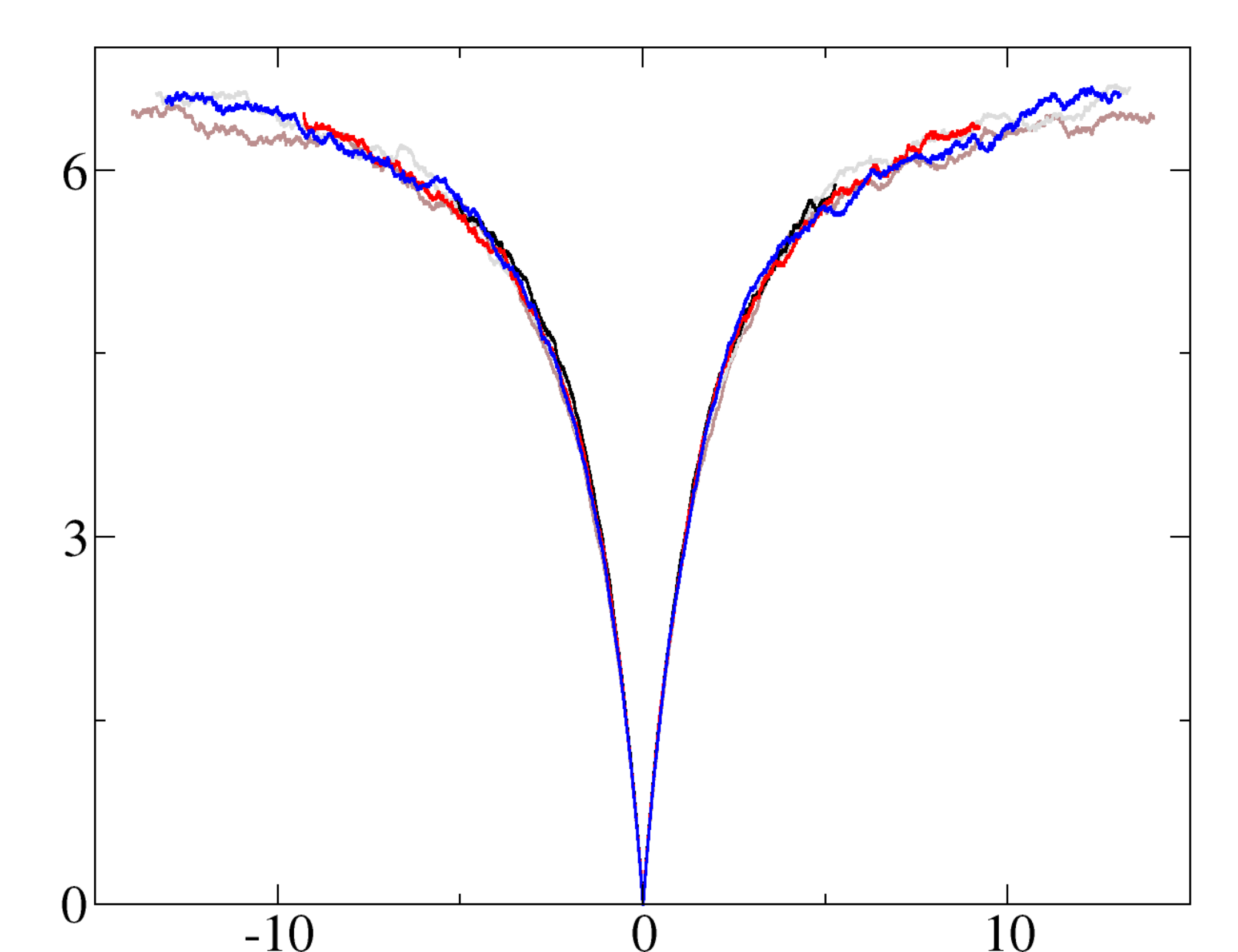}}
    \put(0.52,.14){\includegraphics[width=.3\columnwidth,clip=true]{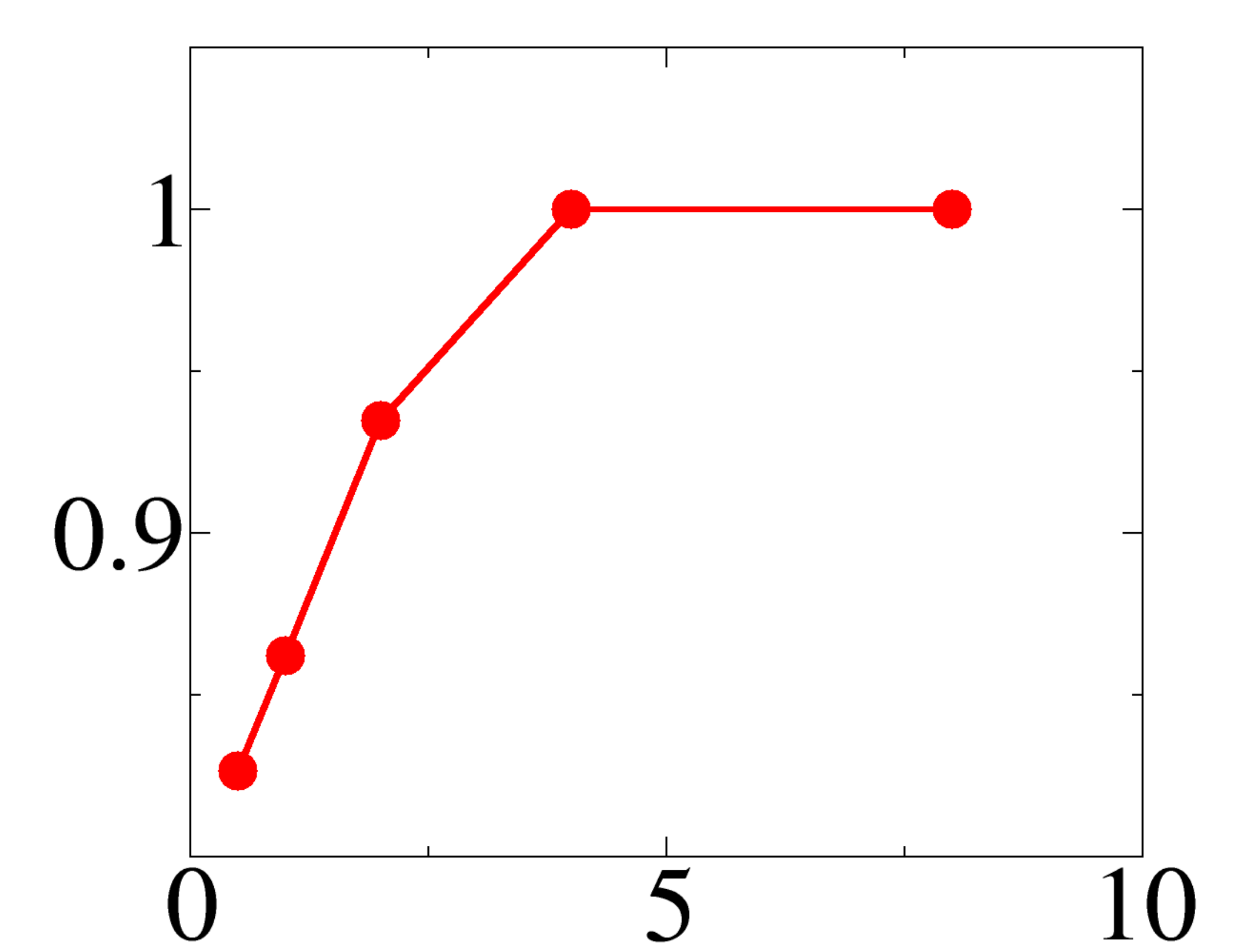}}
    \put(-0.03,0.15){\color[rgb]{0,0,0}\makebox(0,0)[lb]{\rotatebox{90}{\smash{$[\widetilde D(T)/D]^{-1}[B(t)]^{-\frac 12}\Cbar(t,y)$}}}}%
    \put(0.4,0){\color[rgb]{0,0,0}\makebox(0,0)[lb]{\smash{$[B(t)]^{-\frac 12}y$}}}%
    \put(0.52,0.185){\color[rgb]{0,0,0}\makebox(0,0)[lb]{\footnotesize{\rotatebox{90}{\smash{$\widetilde D(T)/D$}}}}}%
    \put(0.68,0.1){\color[rgb]{0,0,0}\makebox(0,0)[lb]{\footnotesize{\smash{$T$}}}}%
  \end{picture}%
  \caption{\label{f:CF-t16384-D}(Color online) Full temperature-independent rescaled
    free-energy fluctuations $\Cbar(t,y)$ for the discrete DP at a
    fixed large time, according to the scaling~\eqref{eq:CF-scal-B-discreteDP}. Same data as in Fig.~\ref{f:CF-t16384}. The
    inset shows the parameter  $\widetilde D$ of~\eqref{eq:CF-scal-B-discreteDP} as a function of $T$.}
\end{figure}
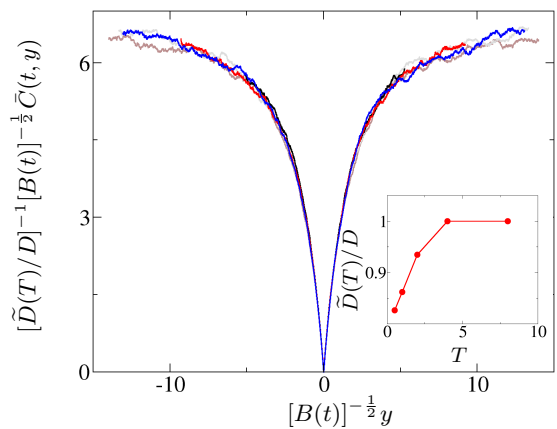

Since the temperature dependence is included in the roughness~$B(t)$,
this scaling relation can also be probed at lower temperatures. This
is done in Fig.~\ref{f:CF-T0.5_1}, were data corresponding to two low
temperatures $T=0.5,1$ and different time scales are collapsed on a
single curve. However, the high- and low-temperature regimes are not
necessarily described by the same rescaling function $\hat C(\bar
y)$. Fig.~\ref{f:CF-t16384} actually shows, for a single value
$t=16384$, that although high- and low-temperature data collapse on a
single curve proper to each regime, there is a crossover between these
two limiting cases.

As suggested by the scaling
relation~\eqref{eq:scaling_Cyt_with-Bt-linear}, the slope of the linear
initial growth of the free-energy correlator should fully account for
the temperature dependence.
As extensively discussed in~\cite{agoritsas_Dtilde_2012},
it can in fact be argued that the
$(D,T)$-dependence of the scaling law~\eqref{eq:scaling_Cyt_with-Bt}
can be absorbed in a single prefactor~$\widetilde D$,
which in the case of the discrete DP model reads
\begin{equation}
 \label{eq:CF-scal-B-discreteDP}
 \Cbar(t,y) = \widetilde D \sqrt{B(t)} \; \hat C^1\Big( \tfrac{y}{\sqrt{B(t)}} \Big),
\end{equation}
where $\widetilde D=\widetilde D (D,T)$ 
and the function $\hat C^1(\bar y)$ is independent of
the parameters $D$ and $T$. 
Indeed, the prefactor $\widetilde D$ in~\eqref{eq:CF-scal-B-discreteDP}
is the slope of the correlator $\Cbar(t,y)$ in the regime
$|y|\lesssim\sqrt{B(t)}$ where $\Cbar(t,y)\propto|y|$.
It is known from the large-time limit at zero $\xi$ that $\widetilde D
= \frac{cD}{T}$, see~\eqref{eq:infinitet_zeroxi_Cbar} (and
also~\eqref{eq:resClinscaling} in the diffusive regime).
In the discrete DP case ($c=T$), $\widetilde D$ is thus expected to be
temperature independent in the high temperature regime, $\widetilde
D=D$, and to behave as $T^{2/3}$ in the low temperature
regime~\cite{agoritsas_Dtilde_2012}. In order to test this, we rescale
the data in Fig.~\ref{f:CF-t16384} onto a single universal master curve as
shown in Fig.~\ref{f:CF-t16384-D}. To this purpose, we fix $\widetilde
D=D=1$ for $T=4,8$ and ensure the collapse of low temperature
correlators $\Cbar(t,y)$ by proposing values for the parameter
$\widetilde D/D$ for $T=0.5,1,2$, which are plotted in the inset of
Fig.~\ref{f:CF-t16384-D}. Although the value of $\widetilde D$ is
decreasing with $T$ we did not observe the $T^{2/3}$ behavior,
possibly because we are not reaching the corresponding low temperature
asymptotic regime for the discrete DP model.
However it is remarkable to observe the crossover towards the
lower-temperature regime, which reflects the influence of the lattice
spacing even though no lengthscale is defined below it.

\begin{figure}
\includegraphics[width = 0.95\columnwidth]{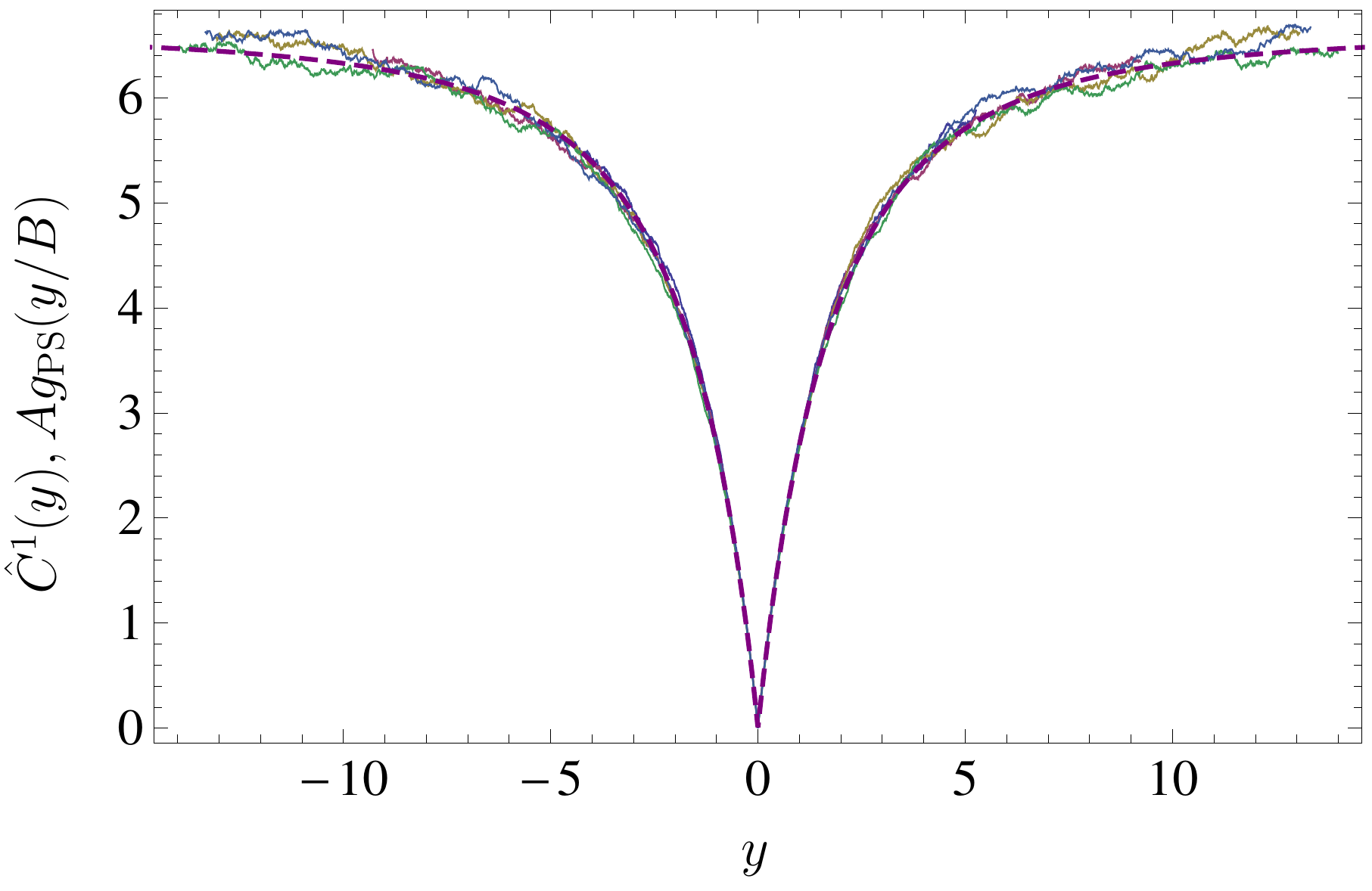}
\caption{(Color online) Comparison between $\hat C^1(y)$ defined in
  Eq.~(\ref{eq:CF-scal-B-discreteDP}), extracted from the numerical
  data shown in Fig.~\ref{f:CF-t16384-D} and $A \, g_{\rm PS}(y/B)$
  (dashed line), where $g_{\rm PS}(y)$ is defined in
  Eq.~(\ref{eq:C-Airy2}) while $A$ and $B$ are two fitting
  parameters (determined to $A\simeq 4.08\pm 0.05$ and $B\simeq 2.07\pm 0.06$). The numerical evaluation of the exact expression for
  $g_{\rm PS}(y)$ was done in
  Refs.~\cite{bornemann_goe_2008,bornemann_fredholm_2010}.}\label{fig:comparison_c_airy2}
\end{figure}
%
%

In summary, we have shown in this section that numerical simulations
of the discrete version of the DP model with the SOS constraint are
fully compatible with the scaling arguments suggesting that
$\sqrt{B(t)}$ is the relevant transverse scale for free-energy
fluctuations. Moreover, we have also shown that all the temperature
dependence can only be accounted for by using, in addition to the
proper transverse length scale $\sqrt{B(t)}$ the scale of linear
free-energy fluctuations at small $y$, given by the function
$\widetilde D$. This universal behavior being established, one would
naturally expect from the result of
Refs.~\cite{prahofer2002,johansson_airy2_2003} in
Eq.~(\ref{eq:scaling-PS}), that this correlator $\hat C^1(y)$ in
Eq.~(\ref{eq:CF-scal-B-discreteDP}) can be expressed in terms of the
mean square displacement of the Airy$_2$ process, ${\cal A}_2(y)$,
namely
\begin{equation}\label{eq:C-Airy2}
\hat C^1(y) = A \, g_{\rm PS}(y/B) \,, \; g_{\rm PS}(y) = \big\langle [{\cal A}_2(y) - {\cal A}_2(0)]^2\big\rangle \;,
\end{equation}

\smallskip
\noindent
where $A$ and $B$ are longitudinal and transverse parameter-dependent
lengthscales. In Fig. \ref{fig:comparison_c_airy2} we test this
relation (\ref{eq:C-Airy2}) by adjusting the parameters $A$ and $B$
using least square fittings and we find indeed a very good collapse of
our numerical data and the exact expression of $g_{\rm PS}(y)$
obtained in Ref. \cite{prahofer2002}. Note that the numerical
evaluation of $g_{\rm PS}(y)$ was done in
Ref.~\cite{bornemann_goe_2008,bornemann_fredholm_2010}, where precise
numerical techniques were developed to compute Fredholm determinants
with high precision.

In the following section we extend this
analysis to lower temperatures and short-range correlations by
studying a continuous version of the DP model.

\section{Numerical simulations at lower temperature: the continuous DP}
\label{sec:numericalresults_continuous}

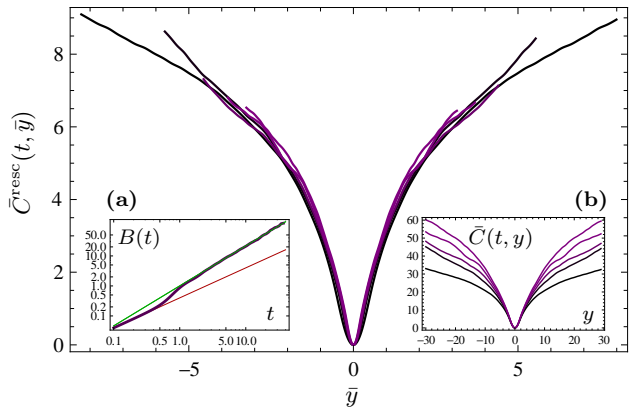
\begin{figure}[!tbp]
\setlength{\unitlength}{.9\columnwidth}
\begin{picture}(1.05,0.7)%
    \put(0,.06){\includegraphics[width=.9\columnwidth]{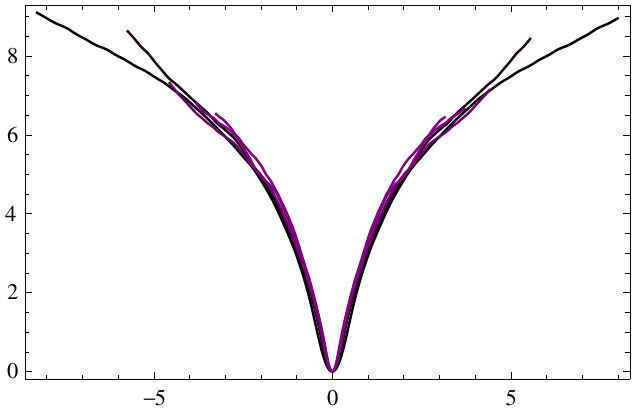}}
    \put(0.61,.12){\includegraphics[width=.32\columnwidth]{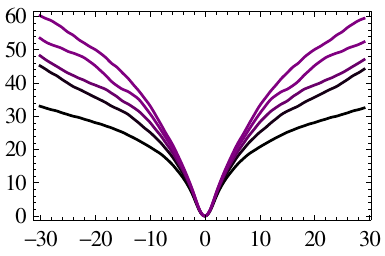}}
    \put(0.065,.124){\includegraphics[width=.32\columnwidth]{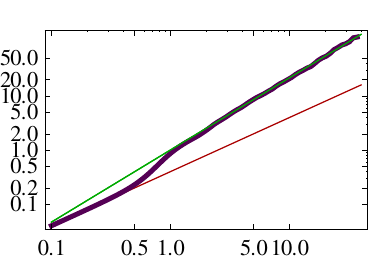}}
    \put(-0.03,0.35){\color[rgb]{0,0,0}\makebox(0,0)[lb]{\rotatebox{90}{\smash{$\Cbar^{\resc}(t,\bar y)$}}}}%
    \put(0.51,0.03){\color[rgb]{0,0,0}\makebox(0,0)[lb]{\smash{$\bar y$}}}%
    \put(.12,0.3){\color[rgb]{0,0,0}\makebox(0,0)[lb]{\rotatebox{0}{\smash{\footnotesize$B(t)$}}}}%
    \put(.375,0.17){\color[rgb]{0,0,0}\makebox(0,0)[lb]{\smash{\footnotesize$t$}}}%
    \put(.72,0.3){\color[rgb]{0,0,0}\makebox(0,0)[lb]{\rotatebox{0}{\smash{\footnotesize$\Cbar(t,y)$}}}}%
    \put(.915,0.17){\color[rgb]{0,0,0}\makebox(0,0)[lb]{\smash{\footnotesize$y$}}}%
    \put(0.1,.365){\color[rgb]{0,0,0}\makebox(0,0)[lb]{\footnotesize\smash{\textbf{(a)}}}}%
    \put(.9 ,.365){\color[rgb]{0,0,0}\makebox(0,0)[lb]{\footnotesize{\smash{\textbf{(b)}}}}}%
  \end{picture}%
  \caption{(Color online) Graph for the continuous DP of the rescaled correlator $\Cbar^{\resc}(t,\bar y)$ defined in~\eqref{eq:defCresc} as a
    function of $\bar y=y/\sqrt{B(t)}$ for different times $t$, at low temperature. Parameters are $c=1$,
    $D=8$, $T=0.4$ and $\xi=2$. Disorder average is performed over 1092 realizations.
    \textbf{Inset (a)}: Roughness $B(t)$ in log-log scale with the thermal (green line) and random manifold (red line) asymptotic regimes.
    \textbf{Inset (b)}: Corresponding bare correlator $\Cbar(t,y)$. Time varies from 35 to 135 by steps of 25 from bottom to top curves.
  }
 \label{fig:Cbarty_T0.4}
\end{figure}

\subsection{Numerical approach}

To probe the implications for the scaling laws of short-range
correlations ($\xi>0$) of the disordered potential $V(t,y)$, we have
simulated a continuous version of the directed polymer.
Directly integrating the partial differential
equation (PDE)~\eqref{eq:eqevolWpolym} for $Z_V(t,y)$ is difficult since,
especially at low temperature, the weight concentrates exponentially in
the most favorable regions for the polymer, preventing the whole space
to be embraced. Taking the logarithm and considering the
PDE~\eqref{eq:eqevolFpolym} for $F_V(t,y)$ is also problematic because
of the singular initial condition $Z_V(0,y)=\delta(y)$.
We took advantage of the STS~\eqref{eq:defFbar} by directly simulating the
PDE for the reduced free-energy $\Fbar_V(t,y)$
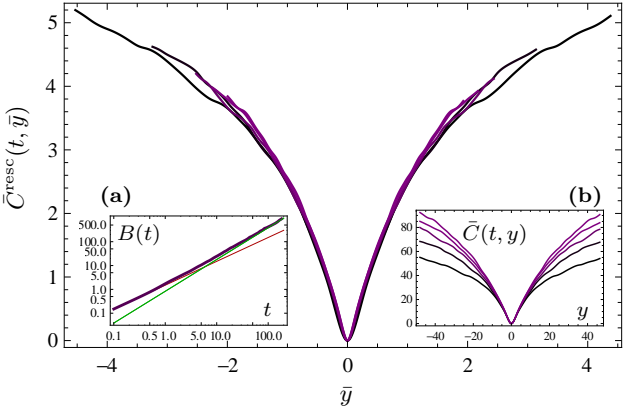
\begin{figure}[!tbp]
\setlength{\unitlength}{.9\columnwidth}
\begin{picture}(1.05,0.72)%
    \put(0,.06){\includegraphics[width=.9\columnwidth]{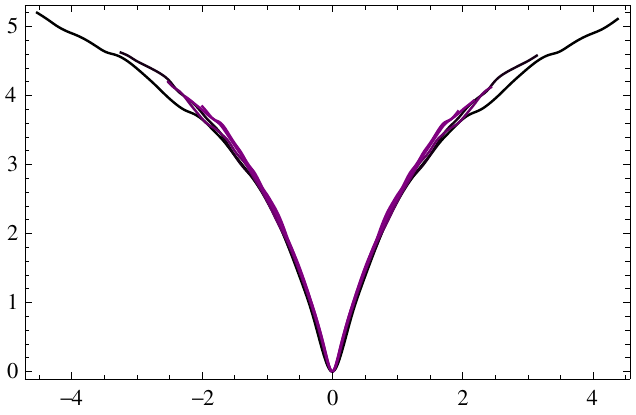}}
    \put(0.61,.12){\includegraphics[width=.32\columnwidth]{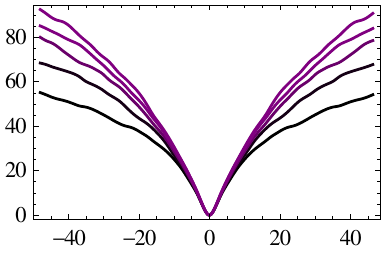}}
    \put(0.065,.128){\includegraphics[width=.32\columnwidth]{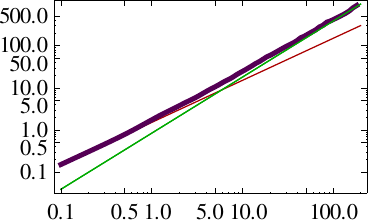}}
    \put(-0.03,0.35){\color[rgb]{0,0,0}\makebox(0,0)[lb]{\rotatebox{90}{\smash{$\Cbar^{\resc}(t,\bar y)$}}}}%
    \put(0.51,0.025){\color[rgb]{0,0,0}\makebox(0,0)[lb]{\smash{$\bar y$}}}%
    \put(.13,0.3){\color[rgb]{0,0,0}\makebox(0,0)[lb]{\rotatebox{0}{\smash{\footnotesize$B(t)$}}}}%
    \put(.38,0.17){\color[rgb]{0,0,0}\makebox(0,0)[lb]{\smash{\footnotesize$t$}}}%
    \put(.72,0.3){\color[rgb]{0,0,0}\makebox(0,0)[lb]{\rotatebox{0}{\smash{\footnotesize$\Cbar(t,y)$}}}}%
    \put(.915,0.17){\color[rgb]{0,0,0}\makebox(0,0)[lb]{\smash{\footnotesize$y$}}}%
    \put(0.1,.365){\color[rgb]{0,0,0}\makebox(0,0)[lb]{\footnotesize\smash{\textbf{(a)}}}}%
    \put(.9 ,.365){\color[rgb]{0,0,0}\makebox(0,0)[lb]{\footnotesize{\smash{\textbf{(b)}}}}}%
  \end{picture}%
  \caption{(Color online) Same graph as Fig.~\ref{fig:Cbarty_T0.4} but at higher temperature $T=1.5$ (disorder average is over 446 realizations).
  }
 \label{fig:Cbarty_T1.5}
\end{figure}
%
%
%
\begin{align}
 \partial_{t} \Fbar_V(t,y) = 
  \frac 1{2\beta c} &\partial^2_{y}\Fbar_V(t,y) -\frac{1}{2c}\big[\partial_{y}\Fbar_V(t,y)\big]^2 
  \nonumber \\
  &\qquad\quad -\frac yt \partial_y\Fbar_V(t,y)  + V(t,y)
  \label{eq:evolFbarty}
\end{align}
whose initial condition is simply $\Fbar_V(0,y)=0$. Another benefit of
considering $\Fbar_V(t,y)$ instead of $F_V(t,y)$ is the removal of the quadratic
contribution $c \frac{y^2}{2t}$ 
which eclipses the disorder contribution in~\eqref{eq:eqevolWpolym} at short time.

The correlated disordered potential $V(t,y)$ is constructed as
follows: independent random variables $V_{i,j}$ are drawn from a
centered Gaussian distribution of variance $D^{\text{grid}}$ on a grid
of coordinates $(t,y)=(i\xi_t,j\xi_y)$. The continuum $V(t,y)$ is
defined as the two-dimensional cubic spline of the $\{V_{i,j}\}$ on the grid. Its
distribution is Gaussian and characterized by its two-point correlator, which can be analytically
computed~\cite{agoritsas_Dtilde_2012} and takes the form
\begin{equation}
  \overline{V(t,y)V(t',y')} = D R_{\xi_t}(t'-t) R_{\xi_y}(y'-y) \;,
\end{equation}
where $R_{\xi}$ is a normalized smooth delta function of width~$\xi$.
The amplitude of disorder is $D=D^{\text{grid}}\xi_t \xi_y$ and our
correlation length of interest in the transverse direction is
$\xi=\xi_y$.
In numerical simulations, we took $\xi_t=1$, $\xi_y=2$, $D^{\text{grid}}=4 $ (and thus $\xi=2$ and $D=8$).  Simulations are
run on a finite window $[-y_\text{m},y_\text{m}]$ in the
transverse direction, outside of which the disorder potential is set
to $0$. Besides, since the equation~\eqref{eq:evolFbarty} is ill-defined
at $t=0$, simulations were run starting at small initial time $t_0$
with thermal initial conditions.
A complete
presentation of the numerical procedure can be found
in Ref.~\cite{agoritsas_Dtilde_2012}.

\subsection{Results}
\label{ssec:result_continuous-DP}

%

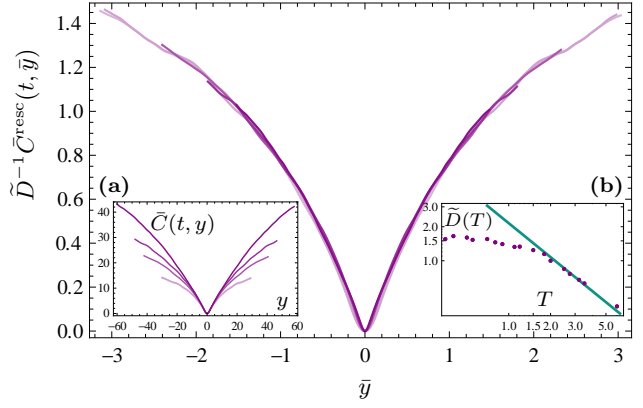
\begin{figure}[!tbp]
\setlength{\unitlength}{.9\columnwidth}
\begin{picture}(1.05,0.7)%
    \put(0,.06){\includegraphics[width=.9\columnwidth]{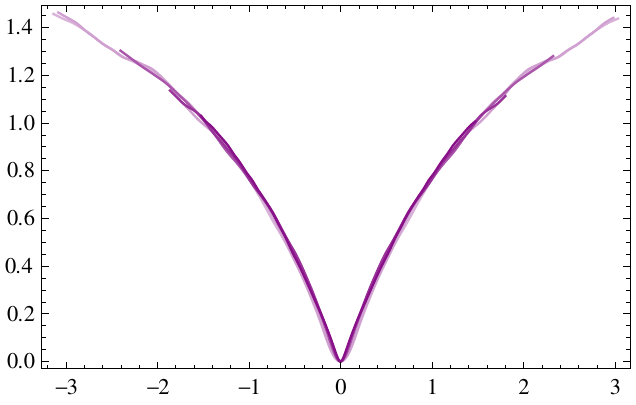}}
    \put(0.635,.125){\includegraphics[width=.315\columnwidth]{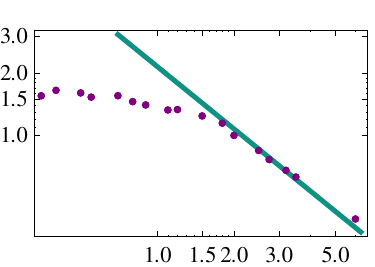}}
    \put(0.08,.125){\includegraphics[width=.315\columnwidth]{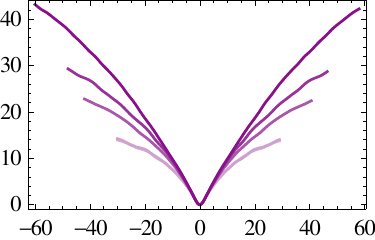}}
    \put(-0.03,0.35){\color[rgb]{0,0,0}\makebox(0,0)[lb]{\rotatebox{90}{\smash{$\widetilde D^{-1}\Cbar^{\resc}(t,\bar y)$}}}}%
    \put(0.525,0.02){\color[rgb]{0,0,0}\makebox(0,0)[lb]{\smash{$\bar y$}}}%
    \put(.17,0.3){\color[rgb]{0,0,0}\makebox(0,0)[lb]{\rotatebox{0}{\smash{\footnotesize$\Cbar(t,y)$}}}}%
    \put(.385,0.17){\color[rgb]{0,0,0}\makebox(0,0)[lb]{\smash{\footnotesize$y$}}}%
    \put(.671,0.307){\color[rgb]{0,0,0}\makebox(0,0)[lb]{\rotatebox{0}{\smash{\scriptsize$\widetilde D(T)$}}}}%
    \put(.83,0.17){\color[rgb]{0,0,0}\makebox(0,0)[lb]{\smash{\footnotesize$T$}}}%
    \put(0.08,.365){\color[rgb]{0,0,0}\makebox(0,0)[lb]{\footnotesize\smash{\textbf{(a)}}}}%
    \put(.92 ,.365){\color[rgb]{0,0,0}\makebox(0,0)[lb]{\footnotesize{\smash{\textbf{(b)}}}}}%
  \end{picture}%
  \caption{(Color online) Plot of the rescaled correlator $\widetilde D^{-1}\Cbar^{\resc}(t,\bar y)$ as a
    function of $\bar y$ at fixed large time $t$ and for different temperatures $T\in\{.4,.55,1.2,1.5,1.8\}$. 
    \textbf{Inset (a)}: corresponding original correlator $\Cbar(t,y)$.
    Lighter color corresponds to lower temperature.
    \textbf{Inset (b)}: $\widetilde D$ as a function of $T$ (up to a numerical factor) in log-log scale. Points are
determined by least-square minimization (see text) with respect to a reference curve at $T=2$ which fixes
$\widetilde D|_{T=2}=1$. 
The line gives the expected slope of the large temperature asymptotics $\widetilde D\propto \frac 1T$ .
  }
 \label{fig:Cbarty_variousT}
\end{figure}
%

For numerical simplicity, the scaling~\eqref{eq:scaling_Cyt_with-Bt}
was tested graphically at fixed~$\xi$, defining a rescaled transverse
coordinate ${\bar y=y/\sqrt{B(t)}}$ and a rescaled correlator
\begin{equation}
  \Cbar^{\text{resc}}(t,\bar y)=
  \frac 1{\sqrt{B(t)}}
  \Cbar_\xi\big( t,\bar y\sqrt{B(t)}\big)
\label{eq:defCresc}
\end{equation}
Indeed testing the full scaling~\eqref{eq:scaling_Cyt_with-Bt} where
$\xi$ is also rescaled by $\sqrt{B(t)}$ would imply to measure
$\Cbar_\xi(t,y)$ for many different values of $\xi$ which was not
numerically accessible [see below for an analytical explanation on the
rescaling $\xi/\sqrt{B(t)}$ of~\eqref{eq:scaling_Cyt_with-Bt}].
Note that in the discrete DP it was sufficient to study the scaling
function of~\eqref{eq:CF-scal-B} without taking~$\xi$ into account (no length below the
lattice spacing, which plays the role of~$\xi$, can be considered).

Low-temperature results are shown on Fig.~\ref{fig:Cbarty_T0.4}:
the curves of $\Cbar_\xi(t,y)$ at different times superimpose
upon the rescaling~\eqref{eq:defCresc}, with a slight discrepancy around
the origin due to the fixed disorder correlation length~$\xi$.
At higher temperature (Fig.~\ref{fig:Cbarty_T1.5}) this scaling
remains valid, the effect of the finite~$\xi$ being as expected less important.
In the evaluation of $\Cbar^{\text{resc}}(t,\bar y)$
from~\eqref{eq:defCresc}, the roughness $B(t)$ is determined from the
numerical results for $\Fbar(t,y)$ using~\eqref{eq:averageOZty} to
evaluate the thermal average $\langle y(t)^2\rangle_V$ at fixed
disorder, averaging afterwards over the realizations of disorder. For
completeness, the graphs of the roughness are given in the inset (a)
of Figs.~\ref{fig:Cbarty_T0.4} and~\ref{fig:Cbarty_T1.5}.

Similarly to the case of the discrete polymer~\eqref{eq:CF-scal-B-discreteDP}, it can be argued that
the $(c,D,T)$-dependence of the scaling
law~\eqref{eq:scaling_Cyt_with-Bt} can be absorbed into a single
prefactor~$\widetilde D$~\cite{agoritsas_Dtilde_2012}
\begin{align}
  \Cbar_\xi(t,y)\  &= \ \widetilde D \sqrt{B(t)}\, \ \hat C^1_{\frac{\xi}{\sqrt{B(t)}}}\Big(\tfrac y{\sqrt{B(t)}}\Big) \;,
 \label{eq:scaling_Cyt_Dtilde}
\end{align}
where the function $\hat C^1_{\bar \xi}(\bar y)$ does not depend on $(c,D,T)$.
By tuning appropriately $\widetilde D$, we show on
Fig.~\ref{fig:Cbarty_variousT} from the numerical evaluation of the
correlator that this is indeed the case.
The function $\widetilde D(T)$ at fixed $c$ and $\xi$ is evaluated by
fixing a reference curve $\Cbar^{\text{resc}}(t,\bar y)$ at
$T=T_\text{ref}$ ($T_\text{ref}=2$ in Fig.~\ref{fig:Cbarty_variousT})
and finding the best  $\widetilde D(T)$ which minimizes the
distance between $\Cbar^{\text{resc}}(t,\bar y)$ and the reference curve
using the least square method. We checked that the result does not depend on 
the choice of $T_\text{ref}$ nor on the choice of the fixed $t$,
within numerical uncertainty.
This method allows a determination of $\widetilde D(T)$ up to a
$T$-independent constant which imposes  $\widetilde D(T_\text{ref})=1$.
We refer the reader to Ref.~\cite{agoritsas_Dtilde_2012} for an in-depth
analytical and numerical study of~$\widetilde D$, regarding
its scaling and physical interpretation.
As displayed in the inset (b) of Fig.~\ref{fig:Cbarty_variousT},
the measured~$\widetilde D$ is compatible with the predicted high $T$ behavior
$\widetilde D=\frac{cD}T$ and a saturation at low~$T$.

We can actually provide a short argument explaining the rescaling of the correlation length $\xi$ by
$\sqrt{B(t)}$ appearing in~\eqref{eq:scaling_Cyt_Dtilde}, that we
could not probe numerically. As the rounding of
$\Cbar(t,y)$ is quadratic in $y$ for small $y$ (see
also~\eqref{eq:Cbarlin_smally_larget}), we expect that $\hat C^1_\xi(\bar y)$
scales as
\begin{align}
  \hat C^1_{\bar \xi}(\bar y)\ {\simeq}\ &\:
\widetilde D \frac{\bar y^2}{{\bar \xi}} \qquad\text{for}\quad |\bar y|\ll ({{\bar \xi}/{\widetilde D}})^{1/2} \;.
\end{align}
The rescaling $\bar\xi=\xi/\sqrt{B(t)}$ in~\eqref{eq:scaling_Cyt_Dtilde} ensures thus 
that, in the large time limit of~\eqref{eq:scaling_Cyt_Dtilde},
$\Cbar_\xi(t,y)$ remains independent of $t$ at small $y$:
\begin{equation}
 \Cbar_\xi(t,y) \xrightarrow[t\to\infty]{y\approx 0}\frac{\widetilde D}{2\xi\sqrt{\pi}}y^2 \;.
\end{equation}
Without the precise rescaling $\xi/\sqrt{B(t)}$ in~\eqref{eq:scaling_Cyt_Dtilde} this expression
would become singular.

Note that because of the rounding at small $y$
the fitting procedure of $\hat C^1(t,y)$ with respect to $g_\text{PS}$, as done
for the discrete DP in Fig.~\ref{fig:comparison_c_airy2}, cannot be performed since the Airy$_2$ process is not adapted to a finite disorder correlation length~$\xi$.

\section{Discussion and conclusion}
\label{sec:conclusion}

In the one-dimensional KPZ universality class,
a suitably centered observable $\Fbar(t,y)$
presents, in the large time limit,
universal fluctuations of order $t^{\frac 13}$ at a transverse
scale $y\sim t^{\frac 23}$,
described by the scaling relation~\eqref{eq:fluct_Fbar_KPZ}.
We have probed numerically a refinement of this scaling law at short
times, for the case of a random environment presenting correlations at
a typical finite lengthscale $\xi$ and at finite temperature, by testing the
scaling relation~\eqref{eq:scaling_Cyt_Dtilde} for the two-point
free-energy correlation function $\Cbar_\xi(t,y)$ defined in~\eqref{eq:defCbar}.
If this scaling extends to the
higher order correlation functions, it will be equivalent to stating that
$\Fbar(t,y)$ scales in distribution as 
\begin{equation}
  \Fbar(t,y) \ {\sim}\ \Big[\widetilde D  \sqrt{B(t)}\Big]^\frac12\: \chi_{\frac \xi{\sqrt{B(t)}}}\big(y/\sqrt{B(t)}\big) \;.
  \label{eq:generic_rescalingDtildexi_forFbar}
\end{equation}

Compared to the large time asymptotics~\eqref{eq:fluct_Fbar_KPZ}, this
relation expresses the fact that to the transverse fluctuations
${\overline{\langle y^2(t)\rangle}=B(t)}$ correspond free-energy
fluctuations of order $\sqrt{B(t)}$, described by the roughness $B(t)$
not only in the RM regime where $B(t)\propto t^{\frac 43}$ but also at
smaller times.
The properly rescaled function $\chi_{\bar\xi}(\bar y)$ would thus be independent of the parameters
$c$, $D$, $T$ and $\xi$, the constant $\widetilde D$ capturing all
the parameters dependence. We have in particular recovered numerically that the
value~\cite{HHF_1985} $\widetilde D=\frac{cD}{T}$ describes correctly
the high-temperature regime where the disorder correlation
length~$\xi$ plays no rôle and correlations take the
form~\eqref{eq:Cofyt_infinite-t}.
Moreover, the distribution of the variable $\chi_{\bar\xi}(\bar y)$ is known at
$\bar\xi=0$ ($\chi_0(y)=\chi(y)$ being then the  ${\text{Airy}}_2$
process~\cite{prahofer2002}) and an interesting open question is thus to
characterize the analogous process at non-zero~$\bar\xi$.
One can conjecture that a suitable characterization is provided by an
appropriate Macdonald process (or a generalization of it) --~already
known~\cite{borodin_macdonald_2011,borodin_free_2012} to yield the
process $\chi(y)$ at $\bar\xi=0$ in some specific limit~-- especially in view
of recent results \cite{oconnell_directed_2012} allowing to represent
Macdonald processes in terms of Brownian motions interacting within a
finite range.
%
%
This could in particular help to characterize the universality of the 
process $\chi_{\bar\xi}(\bar y)$, \emph{i.e.} to determine
how many details of the correlator $R_\xi(y)$
of the microscopic disorder~\eqref{eq:VVcorrelator_roundeddelta}
are left in~$\chi_{\bar\xi}(\bar y)$.

A point of particular interest is the scaling of fluctuations with respect
to temperature.
Regarding the roughness $B(t)$ it has been
shown~\cite{nattermann_interface_1988,bustingorry2010,agoritsas_disordered_ECRYS} that, above a characteristic
temperature~$T_c=(\xi c D)^{\frac
  13}$, $B(t)$ can be rescaled as
\begin{align}
 B(t;c,D,T) &= \xi_{\text{th}}^2 \hat B(t/t_*)  \;,
 \\ \text{with~}\ &
 \xi_{\text{th}}=\frac{T^3}{cD}\;, 
 t_*=\frac{T^5}{cD^2} \;,
\end{align}
while below $T_c$ such a rescaling does not hold.
Nevertheless, our results on the free-energy correlator $\Cbar(t,y)$
indicate that the extended scaling law~\eqref{eq:scaling_Cyt_Dtilde}
holds below and above $T_c$ at large enough times, the dependence
in the parameters $(c,D,T,\xi)$ being then gathered in 
a single prefactor $\tilde D$.
If the roughness exponent $\zeta_\text{RM}$ is not expected to change
below $T_c$, however the prefactor  $\hat A$ of the roughness in $B(t)\sim \hat A t^{2\zeta_\text{RM}}$
can be modified by $\xi$.
This question is relevant in particular to determine the precise
dependence of the crossover timescales of one-dimensional interfaces
described by the DP coordinate
$y(t)$~\cite{agoritsas2010,agoritsas_disordered_ECRYS}.
We refer the reader to~\cite{agoritsas_Dtilde_2012} for numerical
and analytical results in that direction.

On the other hand, recent experiments on liquid crystals
\cite{takeuchi_universal_2010,takeuchi_growing_2011,takeuchi_evidence_2012}
(see also~\cite{miettinen_experimental_2005} for a burning front
experiment) have demonstrated that the height fluctuations of a growth interface are
correctly described by the (centered) free-energy $\Fbar(t,y)$  itself. In particular
it was found that different Airy
processes of the KPZ universality class successfully account for the
observed height distribution (Airy$_1$ or Airy$_2$ depending on the
geometry of the initial condition).
The authors of those studies were in particular able to measure
the correlator $\Cbar(t,y)$ with high precision.
Given the existence of finite disorder correlation length $\xi$ in 
experiments, it would be interesting to study the equivalent
of the low- and high-temperature regimes when tuning the system
parameters which plays the role of $T$ in our description.
In the notation of~\cite{takeuchi_evidence_2012}, 
the relation between the amplitude $\Gamma$
of $\chi(y)$ (linked to our $\widetilde D$ in~\eqref{eq:generic_rescalingDtildexi_forFbar}
by $\Gamma=\widetilde D^2/c$)
and $\lambda$ and $\nu$ in the KPZ equation (linked to our parameters by $c=\tfrac 1\lambda$ and $T=\tfrac\nu\lambda$)
would change from low~$\nu$ to high~$\nu$.
Precisely, defining the characteristic value ${\nu_c=(\xi\lambda^2D)^{\frac 13}}$,
we predict a crossover from the (observable)
dependence $\Gamma\sim D^2 \lambda \nu^{-2}$
at high $\nu$ ($\nu\gg\nu_c$) to a saturated regime
$\Gamma\sim D^2 \lambda \nu_c^{-2}$ at low $\nu$ ($\nu\ll\nu_c$).
The influence of~$\xi$ could thus be probed by measuring
the power-law dependence of~$\Gamma$ in the parameter $\lambda$
displaying a non-trivial dependence $\Gamma\sim\lambda^{-\frac 13}$
for $\nu\ll\nu_c$.

\acknowledgments 

We would like to thank Christophe Berthod for his help on the Mafalda
cluster at the University of Geneva, where part of the simulations were run, and Folkmar
Bornemann for sharing with us his data for the two-point correlation
of the Airy$_2$ process.
E.A. and T.G. acknowledge support by the Swiss National Science
Foundation under MaNEP and Division II.
V.L. was financially supported by ANR SHEPI.
G.S. acknowledges support by ANR grant 2011-BS04-013-01 WALKMAT.
S.B. and G.S. acknowledge partial support by the France-Argentina MINCYT-ECOS
A08E03.
S.B. is partially supported by CONICET grant PIP11220090100051.


\appendix

\section{Different formulations of the free-energy correlator }
\label{app:defsCpropsC}

We determine in this appendix some properties of the free-energy
correlator $\Cbar_\xi(t,y)$ defined in~\eqref{eq:defCbar}.
Defined by the decomposition~\eqref{eq:defFbar} of the free-energy,
the reduced free-energy $\Fbar(t,y)$  is statistically invariant by translation along $y$,
which implies that all its moments  are invariant by translation along $y$ at fixed time: 
\begin{align}
  \overline{\Fbar_V(t,y_1+Y)\ldots\bar
    F_V(t,y_n+Y)}=\overline{\Fbar_V(t,y_1)\ldots\Fbar_V(t,y_n)}\:.
\end{align}
In particular, $\overline{\Fbar(t,y)}$ is independent of $y$,
so that $\Cbar_\xi(t,y)$ is also the connected correlator of $F_V(t,y)$.
Indeed, starting from~\eqref{eq:defFbar}
\begin{align}
  \overline{F_V(t,y)}&= \overline{c \frac {y^2}{2t}+ \frac T2\log \frac{2\pi Tt}c+ \Fbar_V(t,y)} \\
  &= c \frac {y^2}{2t}+ \text{cte}(t)
\end{align}
thus
\begin{align}
\Fbar_V(t,y)&= F_V(t,y)-c \frac {y^2}{2t}- \frac T2\log \frac{2\pi Tt}c  \\
&=  F_V(t,y) -\overline{F_V(t,y)}+ \text{cte}(t)
\end{align}
and hence from the definition~\eqref{eq:defCbar}, and dropping the index $V$ for simplicity
\begin{small}
  \begin{align}
    \Cbar_\xi(t,y'-y)&=
    \overline{\Big\{\big[F(t,y')-\overline{F(t,y')}\big]-\big[
      F(t,y)-\overline{F(t,y)}\big]\Big\}^2}
    \\
    &= \overline{\big[F(t,y')- F(t,y)\big]^2} -\Big[\overline{F(t,y')-
      F(t,y)}\Big]^2\,,
    \label{eq:barCconnected}
  \end{align}
\end{small}

\noindent as announced. The fact that this expression depends only on $y'-y$
is not obvious and arises from the STS through~\eqref{eq:defFbar}.
See Ref.~\cite{agoritsas_Dtilde_2012} for a study of the time dependence
of the quantity $\overline{\Fbar_V(t,y)}$ which is independent of~$y$.

Note that the second derivative of $\Cbar_\xi(t,y)$ with respect to $y$
is directly related to the correlator of the ``phase'' $\eta_V(t,y)\equiv \partial_y
\Fbar_V(t,y)$.  Indeed, using $\Fbar_V(t,y)-\bar
F_V(t,0)=\int_0^{y}dy_1 \eta_V(t,y_1) $ one has
\begin{align}
  \partial_y^2 \Cbar_\xi(t,y)
&= \partial_y^2\overline{\big[\Fbar(t,y)-\Fbar(t,0)\big]^2}
\\
  &= \partial_y^2  \int_0^{y}dy_1\int_0^{y}dy_2 \:\overline {\eta_V(t,y_1) \eta_V(t,y_2)}
\\
  &= \partial_y^2  \int_0^{y}dy_1\int_0^{y}dy_2 \:\Rbar(t,y_2-y_1) 
\\
  &= \Rbar(t,y)+\Rbar(t,-y)  
\\
  \partial_y^2  \Cbar_\xi(t,y) &= 2\Rbar(t,y)\:,
  \label{eq:linkCxiR}
\end{align}
where the correlator of $\eta_V$ is denoted by
\begin{equation}
  \Rbar(t,y_2-y_1) = \overline {\eta_V(t,y_1) \eta_V(t,y_2)}\:.
  \label{eq:defRty}
\end{equation}

\section{Large $y$ behavior of the correlator~$\Cbar_\xi(t,y)$}
\label{app:constantClargey}

In this appendix, we show that, as discussed in section~\ref{ssec:generalizedscaling},
the correlator $\Cbar(t,y)$ presents `wings' at all finite times,
\emph{i.e.} goes to a finite constant $\lim_{|y|\to\infty} \Cbar(t,y)$ at large $|y|$ for $t<\infty$.

To this aim, let us first show that the integral ${\mathcal N(t)=\int dy\,\Rbar(t,y)}$
of the two-point correlator of $\eta_V(t,y)$, defined
in~\eqref{eq:defRty}, is a conserved quantity.
From~\eqref{eq:eqevolFpolym}, the evolution equation for the reduced
free-energy $\Fbar_V(t,y)$ of the directed polymer, defined
in~\eqref{eq:defFbar}, is given by~\eqref{eq:evolFbarty}
\begin{align}
 \partial_{t} \Fbar_V(t,y) = 
  \frac 1{2\beta c} &\partial^2_{y}\Fbar_V(t,y) -\frac{1}{2c}\big[\partial_{y}\Fbar_V(t,y)\big]^2 
  \nonumber \\
  &\qquad\quad -\frac yt \partial_y\Fbar_V(t,y)  + V(t,y)
  \label{eq:evolFbar}
\end{align}
so that the evolution equation of 
$\eta_V(t,y)= \partial_y \Fbar_V(t,y)$ also includes an explicit time dependence:
\begin{align}
 \partial_{t} \eta_V(t,y) = 
  &\frac 1{2\beta c} \partial^2_{y}\eta_V(t,y)  -\frac{1}{c} \eta_V(t,y)\partial_{y}\eta_V(t,y) \nonumber \\ & \
  -\frac 1t \eta_V(t,y) - \frac yt \partial_{y}\eta_V(t,y)  + \partial_yV(t,y)\:.
  \label{eq:eq-evol-eta}
\end{align}
One can now compute $\partial_t\int dy\,\Rbar(t,y)$. Taking advantage
of the invariance of the disorder distribution through the symmetry by
reflection $y\mapsto -y$, and of the statistical invariance of
$\eta_V(t,y)$ by translation along  $y$, one has, dropping the index $V$ for
simplicity
\begin{widetext}
  \begin{small}
    \begin{align}
      \partial_t\int dy\,\Rbar(t,y) &= \partial_t\int
      dy\,\overline{\eta(t,y)\eta(t,0)}
      \\
      &=\int dy\bigg\{ \overline{\eta(t,y)\Big[ \frac 1{2\beta
          c} \partial^2_{y}\eta(t,0) -\frac{1}{c}
        \eta(t,0)\partial_{y}\eta(t,0) -\frac 1t \eta(t,0)
        + \partial_yV(t,0)\Big] }
      \\
      &\qquad\qquad\quad + \overline{\eta(t,0)\Big[ \frac 1{2\beta
          c} \partial^2_{y}\eta(t,y) -\frac{1}{c}
        \eta(t,y)\partial_{y}\eta(t,y) \underbrace{-\frac 1t \eta(t,y)
          - \frac
          yt \partial_{y}\eta(t,y)}_{\text{\footnotesize$=-\frac
            1t \partial_y[y\eta(t,y)]$}}
        + \partial_yV(t,y)\Big]}\bigg\}
      \nonumber \\
      &= -\frac 1t \int dy \:\overline{\eta(t,y)\eta(t,0)} + \frac 1{\beta c}
      \overline{\eta(t,0)\big[\partial_y\eta(t,y)-\partial_y\eta(t,0)\big]_{y=-\infty}^{y=+\infty}} 
       \nonumber
      \\
      &\quad\qquad\qquad -\frac
      1{c}\overline{\eta(t,0)\big[\eta(t,y)^2\big]_{y=-\infty}^{y=+\infty}}
      -\frac 1{t}\overline{\eta(t,0)\big[
        y\eta(t,y)\big]_{y=-\infty}^{y=+\infty}}
      \\
\Longleftrightarrow \   \partial_t\int dy\,\Rbar(t,y) &= -\frac 1t \int dy\,\Rbar(t,y)\:.
    \end{align}
  \end{small}
\end{widetext}
This shows that $\mathcal N(t)$
verifies the equation $\partial_t\mathcal N(t)=-\frac 1t \mathcal
N(t)$ whose solution reads $\mathcal N(t)= \frac{C_1}{t}$.  The
constant $C_1$ is equal to $0$ thanks to the initial condition
$\mathcal N(0)=0$ (the initial condition for $\Fbar$ is
$\Fbar_V(0,y)=0$, see part~\ref{ssec:generalizedscaling}, and thus $\eta_V(0,y)=0$).  This
yields that $\mathcal N(t)$ is constant in time and equal to $0$:
\begin{equation}
  \mathcal N(t)=\int dy\,\Rbar(t,y) = 0\,.
\end{equation}
This equality at all finite times $t$ is equivalent to the existence
of wings in the correlator $\Cbar_\xi(t,y)$ at finite
times. Indeed, since $2 \int_0^y dy'\, \Rbar(t,y')= \partial_y \bar
C_\xi(t,y)$ [see~\eqref{eq:linkCxiR}], this equality also writes
\begin{equation}
\lim_{y\to\infty} \partial_y \Cbar_\xi(t,y)=0\,,
\end{equation}
and this corresponds to the plateau at large $|y|$ of
Fig.~\ref{fig:wingedcorrelator}.  This result is already known to hold
at large finite time and for a delta-correlated disorder
($\xi=0$)~\cite{prahofer2002}. In this appendix we thus have
generalized this result at all times, and to a disorder correlator
with short-range correlations.
%

\section{Short-time dynamics (diffusive scaling)}
\label{sec:shorttime}

In this appendix, we study in the short-time regime
and at finite $\xi$ the scaling behavior of the two-point
correlators $\Cbar_\xi(t,y)$ and $\Rbar_\xi(t,y)$ of $\Fbar_V(t,y)$ and $\eta_V(t,y)=\partial_y\Fbar_V(t,y)$,
defined respectively
in~\eqref{eq:defCbar} and~\eqref{eq:defRty}. 

The polymer is pinned in $y=0$ at time
$t=0$ so that the initial condition  translates for the reduced free-energy into $\bar
F_V(0,y)=0$. Let us thus assume that $\bar
F_V(0,y)$ remains small at short times so that in this regime the non-linear term
of~\eqref{eq:evolFbar} remains small and negligible compared to the
linear terms:
\begin{equation}
 \partial_{t} \Fbar_V(t,y) \simeq
  \frac T{2 c} \partial^2_{y}\Fbar_V(t,y) 
  -\frac yt \partial_y\Fbar_V(t,y)  + V(t,y)\:.
  \label{eq:evolFlinear}
\end{equation}
This equation is linear and can thus be solved directly. Before doing
so, note that for an uncorrelated disorder ($\xi=0$), the 
corresponding steady-state distribution of $\Fbar$ is the
same~\cite{HHF_1985} as in the non-linearized one~\eqref{eq:steadystateKPZ}
(remark from~\eqref{eq:defFbar} that $F$ and $\Fbar$ share the same
distribution at infinite time).
In particular, denoting $\Cbar^\lin_\xi(t,y)$ (resp.
$\Rbar^\lin_\xi(t,y)$) the same correlator as~\eqref{eq:defCbar}
(resp.~\eqref{eq:defRty}) but for $\Fbar_V$ solution of the
linearized equation~\eqref{eq:evolFlinear}, one also has
\begin{align}
  \Cbar^\lin_\xi(t,y)|_{\xi=0}&\xrightarrow[t\to\infty]{}\frac{cD}T|y| \label{eq:Cbarlinasympt}
\\
  \Rbar_\xi^\lin(t,y)|_{\xi=0}&\xrightarrow[t\to\infty]{}\frac{cD}T \delta(y) \,,\label{eq:Rbarlinasympt}
\end{align}
where $\Rbar^\lin_\xi(t,y)=\frac 12 \partial_y^2\Cbar^\lin_\xi(t,y)$ as in~\eqref{eq:linkCxiR}.
Although the solution of the linearized evolution
equation~\eqref{eq:evolFlinear} is valid only at short time compared
to the complete evolution~\eqref{eq:evolFbar}, it is thus instructive to
study its behavior at all times.

In what follows, we determine the finite $t$ and finite $\xi$
equivalents of the correlators~(\ref{eq:Cbarlinasympt},\ref{eq:Rbarlinasympt}).
To get rid of the term $\frac yt \partial_y\Fbar_V$ in~\eqref{eq:evolFlinear} we set
\begin{equation}
  \Fbar_V(t,y) = \sqrt{t} e^{\beta \frac{c y^2}{2t}}\hat F_V(t,y)\:.
\end{equation}
The evolution of $\hat F_V$ is then
\begin{equation}
 \partial_{t} \hat F_V(t,y) =
  \frac 1{2\beta c} \partial^2_{y}\hat F_V(t,y) 
  + \underbrace{\frac{1}{\sqrt{t}} e^{ \frac{-\beta c y^2}{2t}} V(t,y)}_{\text{\normalsize{$\equiv \hat V(t,y)$}}}
 \label{eq:evolFtilde}
\end{equation}
Besides, in absence of disorder ($\hat V=0$), 
\begin{equation}
  G(t,y)= \sqrt{\frac{\beta c}{2\pi t}} e^{-\frac{\beta c y^2}{2t}} \theta(t)
\end{equation}
is a Green function of the equation for $\hat F_V$, that is:
\begin{equation}
  \big[\partial_{t} -\frac 1{2\beta c} \partial^2_{y}\big] G(t-t',y-y') = \delta(t-t')\delta(y-y')\:.
\end{equation}
We denote by $\theta(t)$ the Heaviside step function.
The solution of~\eqref{eq:evolFtilde} is hence
\begin{equation}
  \hat F(t,y)=\int_0^{+\infty} \!\!\! dt' \int\!\! dy' G(t-t',y-y')\hat V(t',y')\:.
  \label{eqref:solFhat}
\end{equation}
Here and in what follows, integrals along direction $y$ run by convention on the real line.

Let us first determine the effect of $\xi$ by computing the value of the peak of the correlator
\begin{equation}
 \begin{split}
 \Rbar^\lin_\xi(t,0)&=
  \frac 12 \partial_y^2\overline{\big[\bar
   F(t,y) - \Fbar(t,y)\big]^2 }\Big|_{y=0}
 \\ &=\overline{\partial_y \bar
   F(t,0)\, \partial_y \Fbar(t,0)}
 \end{split}
\end{equation}
which is a matter of Gaussian integration.
One has $\partial_y \Fbar(t,0)=\sqrt{t}\partial_y\hat F(t,0)$ and
\begin{align}
  \partial_y\hat F(t,0) =&\int_0^{+\infty}\!\!\!\!\!\!\!\!dt_1\!\! \int\!\! dy_1\:  \partial_yG(t-t_1,-y_1)\hat
  V(t_1,y_1)
  \\
  =&\frac 1{\sqrt{2\pi}}\int_0^t\!\!dt_1\!\! \int\!\! dy_1
  \Big(\frac{\beta c}{t-t_1}\Big)^{\frac 32} y_1 
  e^{-\frac{\beta c y_1^2}{2(t-t_1)}} 
  \nonumber \\ & \qquad\qquad\quad\qquad\quad
  \times \frac {e^{ \frac{-\beta c
        y_1^2}{2t_1}}}{\sqrt{t_1}} V(t_1,y_1) 
\end{align}
so that putting everything together, and considering for the disorder $V$
the Gaussian correlator $R^\text{Gauss}_{\xi}(y)$ defined in~\eqref{eq:defRtyGauss}, one gets
\begin{widetext}
\begin{align}
  \Rbar^\lin_\xi(t,0)
  &= t\: \overline{\partial_y \hat F(t,0) \partial_y \hat F(t,0)} 
\\[-8mm]
&= t\:\frac {(\beta c)^3}{2\pi}
  \int\!\! dy_1dy_2\int_0^t\!\!dt_1dt_2 
  \frac{y_1y_2e^{-\frac{\beta c}2\big(\frac{y_1^2}{t-t_1}+\frac{y_2^2}{t-t_2}\big)}}{(t-t_1)^{\frac 32}(t-t_2)^{\frac 32}}
  \frac{e^{-\frac{\beta c}2\big(\frac{y_1^2}{t_1}+\frac{y_2^2}{t_2}\big)}}{\sqrt{t_1t_2}}
  \overbrace{\overline{V(y_1,t_1) V(y_2,t_2)}}^{\frac{D}{2\sqrt\pi\xi}e^{-\frac{(y_2-y_1)^2}{4\xi^2}}\delta(t_1-t_2)}
\\
&= t\: \frac {(\beta c)^3}{4\pi^{\frac 32}\xi} D
  \int dy_1dy_2\int_0^tdt_1 \frac{y_1y_2}{(t-t_1)^{3}t_1}\exp
  \bigg[{-\frac{\beta c}2\bigg(\frac{1}{t-t_1}+\frac 1{t_1}\bigg)(y_1^2+y_2^2)-\frac{(y_2-y_1)^2}{4\xi^2}}\bigg]\:.
\label{eq:Rlinximulticolumngaussianintegrals}
\end{align}
\end{widetext}
The quadratic form in the last exponential writes ${-\frac 12 \vec y\:^T\! A\,\vec y}$ with $\vec y = (y_1,y_2)$ and
$A$ the matrix
\begin{small}
  \begin{equation}
    A =
    \begin{pmatrix}
      \beta c \left(\frac{1}{t_1}+\frac{1}{t-t_1}\right)  +\frac{1}{2 \xi ^2} & -\frac{1}{2 \xi ^2} \\
      -\frac{1}{2 \xi ^2} & \beta c
      \left(\frac{1}{t_1}+\frac{1}{t-t_1}\right) +\frac{1}{2 \xi ^2}
    \end{pmatrix}\:.
  \end{equation}
\end{small}
The result of the Gaussian integral with respect to $\vec y$ is 
\begin{small}
  \begin{equation}
    \big(A^{-1}\big)_{12}\sqrt{\det\frac{2\pi}A} =
    \frac{\pi  \xi  t_1^3 (t-t_1)^3}{\big\{\beta  c t \left[\beta  c \xi ^2 t+t_1 (t-t_1)\right]\big\}^{3/2}}
  \end{equation}
\end{small}
and finally~\eqref{eq:Rlinximulticolumngaussianintegrals} becomes
\begin{small}
  \begin{align}
    \Rbar^\lin_\xi(t,0) &= t \frac {(\beta c)^3}{4\pi^{\frac 32}\xi} D
    \int_0^t\!\!dt_1 \frac{\pi \xi t_1^2}{\big\{\beta c t \left[\beta c \xi
        ^2 t+t_1 (t-t_1)\right]\big\}^{3/2}}
    \\
    = \beta &c D\,\frac{t(t+2\beta c\xi^2)-\xi\sqrt{\beta c
        t}(t+4\beta c\xi^2)
      \operatorname{arccot}\big(2\xi\sqrt{\frac{\beta
          c}t}\big)}{2\sqrt{\pi}t\xi(t+4\beta c\xi^2)}
  \end{align}
\end{small}

We see on one hand that the infinite time limit reads
\begin{equation}
  \lim_{t\to\infty}\Rbar^\lin_\xi(t,0) = \frac{cD}{2T\xi\sqrt{\pi}}
\end{equation}
which diverges indeed as $\xi\to 0$. The effect of $\xi$ is to
regularize the $\xi=0$ result~\eqref{eq:Rbarlinasympt} around $y=0$.
The factor~$\frac1{2\sqrt{\pi}}$ directly arises from the Gaussian
correlator~\eqref{eq:defRtyGauss} as will become clear below.
On the other hand, the short-time behavior writes
\begin{equation}
  \Rbar^\lin(t,0) = \frac{Dt}{12\xi^2\sqrt{\pi}}+O(t^{3/2})\:.
\end{equation}
As for the correlator $\Cbar^\lin_\xi(t,y)$ of the reduced free-energy, this yields
from~\eqref{eq:linkCxiR} the small $y$ expansions
at large and short times
\begin{align}
  \Cbar^\lin_\xi(t,y)\xrightarrow[t\to\infty]{}&\ \frac{cD}{2T\xi\sqrt{\pi}}y^2 \quad\text{for~} y\ll \sqrt{\frac{T\xi}{cD}}
  \label{eq:Cbarlin_smally_larget}
\\
  \Cbar^\lin_\xi(t,y)\stackrel{(t\to 0)}{\approx}&\ \frac{Dt}{12\xi^2\sqrt{\pi}}y^2 \quad\text{for~} y\ll \sqrt{\frac{\xi^2}{Dt}}
\end{align}

The full $(t,y)$-scaling of the correlator $\Cbar^\lin_\xi(t,y)$ can be determined
using the same approach as exposed above, starting from the solution~\eqref{eqref:solFhat} for~$\hat F_V(t,y)$. One finds
  \begin{equation}
    \Cbar^\lin_\xi(t,y)=
    D\int_0^t \!\!dt_1
    \frac{1-\exp \left(-\frac{\beta  c t_1^2 y^2}{4 t \left(\beta  c \xi ^2 t+{t_1} (t-{t_1})\right)}\right)}{
      \sqrt{\pi }
      \sqrt{\frac{{t_1} (t-{t_1})}{\beta  c t}+\xi ^2}}\:.
    \label{eq:resCbarlin}
  \end{equation}
To reveal the scaling of this expression, one performs the change of variable $t_1=t\tau$ in the integral
  \begin{align}
    \Cbar^\lin_\xi(t,y)&= \frac{cD}T \sqrt{\frac t{\beta c}} \int_0^1\!\!\! d\tau \frac{
      1-\exp \big[\frac{-\beta c y^2/t }{4 (\xi ^2 \beta c /t
          +\tau (1-\tau))}\big]}{ \sqrt{\pi } \sqrt{\tau
        (1-\tau)+\xi ^2\beta c/t}}
\nonumber    \\
    &= \sqrt{\frac{t}{\beta c}}\: \Cbar_{\xi\sqrt{{\beta c}/{t}}}\Big(1,y
    \sqrt{{\beta c}/{t}}\Big)
  \end{align}
which takes the scaling form, with the thermal roughness $B_\therm(t)=\frac t{\beta c}=\frac{Tt}{c}$
\begin{align}
  \Cbar^\lin_\xi(t,y)\  &= \ \frac{cD}T \: \sqrt{B_\therm(t)}\, \ \hat C_{\frac{\xi}{\sqrt{B_\therm(t)}}}\Big(\tfrac y{\sqrt{B_\therm(t)}}\Big)
  \label{eq:resClinscaling}
\end{align}
where the scaling function is 
\begin{align}
  \hat C_{\bar\xi}(\bar y)&=\Cbar_{\bar\xi}(t=1,\bar y)|_{c=\beta=D=1}
\\ &=
  \int_0^1\!\!\! d\tau \frac{
      1-\exp \big[-\frac{\bar y^2 }{4 (\bar\xi ^2  +\tau (1-\tau))}\big]}{ \sqrt{\pi } \sqrt{\tau
        (1-\tau)+\bar\xi ^2}}\:.
 \label{eq:reshatCscalingfunction}
\end{align}
The relation~\eqref{eq:resClinscaling} describes a scaling form of the free-energy correlation, with
the roughness $B_\therm(t)$ of the thermal regime. We discuss in
sections~\ref{ssec:result_discrete-DP}
and~\ref{ssec:result_continuous-DP} an extension of its validity to the
large time regime of the DP, with the full roughness $B(t)$ instead of  $B_\therm(t)$.

%

\section{Solution of the winged DP toy model in the Gaussian Variational Method approximation}
\label{app:GVMtoywings}

As introduced in Ref.~\cite{mezard_manifolds_1992,goldschmidt_manifolds_1993},
what we call generically a DP `toy model' is essentially based on the assumption that the reduced free-energy $\bar{F}_V(t,y)$ of the polymer end-point (the `effective potential' it sees) has a Gaussian distribution, \textit{i.e.} that it is fully determined by its mean value and its two-point correlator $\bar{C}^{\text{toy}}(t,y)$.

Following the scheme used in Ref.~\cite{agoritsas2010}, we compute in this appendix the Gaussian-Variational-Method (GVM) approximation of a DP toy model with saturation `wings' appearing at ${|y| \gtrsim \ell_t}$ at time $t$.
We check specifically that by consistency they should appear at ${\ell_t \sim \sqrt{B(t)}}$ at asymptotically large lengthscales, as asserted in Sec.~\ref{ssec:toymodelwings}.

\subsection{Replicae}

The replica approach allows to determine the statistical average of an observable $\mathcal O(t,y)$ 
of the DP toy model as defined in paragraph~\ref{ssec:toymodelwings} from the formal expression
\begin{align}
    \overline{\langle\mathcal O(t,y)\rangle}& = 
    \lim_{n\to 0}
    \int_{\mathbb{R}^n} \!\! dy_1\ldots dy_n  \mathcal O (y_1) 
    \overline{~e^{-\beta\sum_{a=1}^n F^{\toy}(t,y_a)}}
\\
    & =\lim_{n\to 0} \int_{\mathbb{R}^n} \!\! dy_1\ldots dy_n \mathcal O (y_1)
      \exp\{-\beta \tilde F(t,\mathbf y)\}
    \label{eq:averageobs-replica}
\end{align}
where the replicated free-energy $\tilde F(t,\mathbf y)$ of $n$ copies
${\mathbf y =(y_1,\ldots,y_n)}$ of the polymer endpoint reads
\begin{equation}
 \tilde F(t,\mathbf y)=
 \underbrace{\frac c{2t}  \sum_{a=1}^{n} y_a^2}_{\tilde{F}_{\text{el}}(t,\mathbf y)}
 +\underbrace{\frac{\beta}4 \sum_{a,b=1}^n \Cbar^\toy(t,y_b-y_a)}_{\tilde{F}_{\text{dis}}(t,\mathbf y)}\:.
\label{eq:Ftilde_toy}
\end{equation}
To obtain this expression, we have used the assumed Gaussianity of the $\Fbar^{\toy}$ distribution, presumably inherited from the random-potential Gaussian distribution.

\subsection{Gaussian Variational Method (GVM)}

There is no known way of determining the exact roughness of the toy model.
The GVM approach~\cite{mezard_manifolds_1992,goldschmidt_manifolds_1993} for the toy model (see~\cite{agoritsas2010}
for a detailed presentation) consists in approximating the replicated free-energy~\eqref{eq:Ftilde_toy} by a trial quadratic free-energy
\begin{equation}
  \tilde{F}_0 (t)=\frac 12\sum_{a,b=1}^n y_a G^{-1}_{ab}(t)y_b
  \label{eq:def-repli-F0}
\end{equation}
parametrized by the ${n \times n}$ hierarchical matrix ${G^{-1}(t)}$:
\begin{equation}
  G^{-1}_{ab}(t)=\frac ct\delta_{ab}-\sigma_{ab}(t)
  \label{eq:def_invG_sigma_ab}
\end{equation}
with the connected part ${G_c^{-1} (t) = \sum_b G^{-1}_{ab} = \frac{c}{t}}$ fixed by the case in absence of disorder.
The corresponding roughness is directly read from the diagonal term of its inverse matrix, combining \eqref{eq:averageobs-replica} and \eqref{eq:def-repli-F0}:
\begin{equation}
 B(t) = T \lim_{n \to 0} G_{aa}(t)\:.
 \label{eq:def-roughness-replica}
\end{equation}

To find the best quadratic approximation of ${\tilde F(t,\mathbf y)}$, the
extremalization conditions read for the pairs $\left\lbrace a,b \right\rbrace$:
\begin{equation}
  \frac{\partial\Fvar }{ \partial G_{ab} (t)} \left[G(t) \leftrightarrow G^{-1}(t) \right] =0
  \label{eq:extremalizationcondition_GVMtoy}
\end{equation}
where the variational \textit{physical} free-energy $\Fvar$ (\textit{i.e.} after averaging over thermal fluctuations) is defined at each time $t$ as
\begin{equation}
  \Fvar=\mathcal F_0+\langle\tilde F - \tilde{F}_0\rangle_0
  =\mathcal F_0+\langle\tilde F_\text{el} - \tilde{F}_0\rangle_0+\langle\tilde F_\text{dis}\rangle_0
  \label{eq:computeFvar-0}
\end{equation}
and the trial physical free-energy is
\begin{equation}
  \mathcal F_0  = -T\log Z_0 = -\frac T2\log\det G + \text{Cte}
  \label{eq:computeFvar-1}
\end{equation}
In those expression, $\langle\cdot\rangle_0$ denotes the average with
respect to the normalized Boltzmann weight $e^{-\beta \tilde{F}_0}/Z_0$ and $Z_0$ is the corresponding
partition function. Let us first compute
  \begin{align}
    \langle\tilde F_\text{el} - \tilde{F}_0\rangle_0 &= \Big\langle\frac c{2t}
    \sum_{a=1}^n y_a^2 - \frac 12\sum_{a,b=1}^n y_a
    G^{-1}_{ab}(t)y_b\Big\rangle_0
    \\
    &= \frac 12 \sum_{ab} \Big[\frac ct
    \delta_{ab}-G^{-1}_{ab}(t)\Big] \langle y_a y_b\rangle_0
    \\
    &= \frac 12 \sum_{ab} \Big[\frac ct
    \delta_{ab}-G^{-1}_{ab} (t)\Big] T G_{ab}(t)
    \\
    &= \frac{cT}{2t}\sum_{a=1}^n G_{aa}(t) + \text{Cte}  \label{eq:computeFvar-2}
  \end{align}
\noindent and using the Fourier transform of the correlator ${\bar{C}^{\text{toy}}(t,y)}$ as defined in \eqref{eq:barCtyFourier}:
\begin{small}
  \begin{align}
    \langle\tilde F_\text{dis}\rangle_0 &= \frac{\beta \widetilde D}{4}
    \sum_{a,b=1}^n \int_{\mathbb{R}} \frac{d \lambda}{2\pi} \,\frac{2}{\lambda^2} \big\langle 1-\cos
    \big[\lambda (y_b-y_a)\big]\big\rangle_0 R_{\xitilde}^\text{toy}(t,\lambda) \\ &=
    \frac{\beta \widetilde D}{2} \sum_{ab}
   \int_{\mathbb{R}} \frac{d \lambda}{2\pi} \,\frac{1-e^{-\frac 12
      \lambda^2 \langle (y_b-y_a)^2\rangle_0 }   }{\lambda^2}  R_{\xitilde}^\text{toy}(t,\lambda)
    \\ &= \frac{\beta \widetilde D}{2} \sum_{ab}
    \int_{\mathbb{R}} \frac{d \lambda}{2\pi} \,\frac{1-e^{-\frac T2
      \lambda^2 (G_{aa}+G_{bb}-2G_{ab})} }{\lambda^2} R_{\xitilde}^\text{toy}(t,\lambda)\:.
    \label{eq:Fdismoy0_toy}
  \end{align}
\end{small}

Gathering the results \eqref{eq:computeFvar-1}-\eqref{eq:computeFvar-2}-\eqref{eq:Fdismoy0_toy} in the definition of $\Fvar (t)$ \eqref{eq:computeFvar-0}, we obtain an explicit expression of ${\Fvar \left[ G(t) \right]}$ and we can apply the extremalization condition \eqref{eq:extremalizationcondition_GVMtoy} for the off-diagonal terms ${a \neq b}$ (setting the usual notation $\widetilde G=G_{aa}$):
\begin{align}
  0 &=
  -\frac T2 G^{-1}_{ab}
  -2\times\frac{\beta \widetilde D}{2}
  \int_{\mathbb{R}} \frac{d \lambda}{2\pi} \,\frac{\lambda^2 T}{\lambda^2}\,
  e^{- \lambda^2 T (\widetilde G-G_{ab})} R_{\xitilde}^\text{toy}(t,\lambda)
\label{eq:determination_extremeq_toy}
\\ &=
  \frac T2 \sigma_{ab} -
  \widetilde D
 \int_{\mathbb{R}} \frac{d \lambda}{2\pi} \, e^{-  \lambda^2 T (\widetilde G-G_{ab})} R_{\xitilde}^\text{toy}(t,\lambda)
\end{align}
since the sum $\sum_{ab}$ in~\eqref{eq:Fdismoy0_toy} contains twice
$G_{ab}$ by symmetry of the matrix, hence the `$2\times$'
in~\eqref{eq:determination_extremeq_toy}.
Thus for a continuous parametrization $\sigma(u)$ (${u \in \left[ 0,1
  \right]}$, see Ref.~\cite{mezard_variational_replica}) of the hierarchical matrix $\sigma_{ab}$, in a full
replica-symmetry-breaking (full-RSB) formulation, one gets the
self-consistent equation at each time $t$:
\begin{equation}
  \sigma(t,u) = 
  2\frac {\widetilde D} T
  \int_{\mathbb{R}} \frac{d \lambda}{2\pi} \, e^{-  \lambda^2 T [\widetilde G (t)-G(t,u)]} R_{\xitilde}^\text{toy}(t,\lambda)
\label{eq:extr_eq_sigma_toymodel}
\end{equation}
where $\big[ \widetilde{G}(t),G(t,u) \big]$ characterizes the matrix $G(t)$ whose matrix inverse has off-diagonal coefficients ${G^{-1} (t,y) = - \sigma (t,u)}$.

\subsection{Formal solution of the equation for $\sigma(u)$}

The algebra of hierarchical matrices in the limit ${n \to 0}$ has been worked out in Ref.~\cite{mezard_variational_replica}, and following the conventions of Ref.~\cite{agoritsas2010} we recall the properties needed thereafter in the derivation of the GVM solution:
\begin{eqnarray}
 & \partial_u\big[\widetilde G-G(u)\big]
 =-\frac{\sigma'(u)}{\big(G_c^{-1}+[\sigma](u)\big)^2} \label{eq:replica-01} 
 & \\
 & \left[ \sigma \right] (u)
 = u \sigma (u) - \int_0^u dv \, \sigma (v) \label{eq:replica-02} 
 & \\
 & \widetilde{G} - G(u)
 = \frac{1}{u}  \frac{1}{G^{-1}_c + [\sigma](u)} - \int^1_u \frac{dv}{v^2}  \frac{1}{G^{-1}_c +[\sigma](v)} \label{eq:replica-04}
 & \\
 & \widetilde{G}
 = \frac{1}{G^{-1}_c} \left( 1 + \int_0^1 \frac{dv}{v^2} \frac{[\sigma](v)}{G^{-1}_c + [\sigma](v)} + \frac{\sigma (0)}{G^{-1}_c} \right) \label{eq:replica-03}
 &
\end{eqnarray}
where the connected term ${G_c^{-1}(t)=c/t}$.
Denoting
\begin{equation}
  \J_k (t)=\int_{\mathbb{R}} \frac{d \lambda}{2\pi} \, \lambda^k e^{-  \lambda^2 T [\widetilde G(t)-G(t,u)]} R_{\xitilde}^\text{toy}(t,\lambda)
  \label{eq:def-Jk-lambda}
\end{equation}
the extremalization equation~\eqref{eq:extr_eq_sigma_toymodel} becomes simply ${\sigma(u)=2\beta\widetilde D \J_0}$.
Differentiating with $\partial_u$ and assuming that ${\sigma'(u) \neq 0}$, we obtain
\begin{equation}
  1=\frac{2\widetilde D}{\big[ G_c^{-1}+[\sigma](u)\big]^2} \J_2\;.
  \label{eq:diffu_GVMtoy}
\end{equation}
Isolating ${G_c^{-1}+[\sigma](u)=(2 \widetilde{D} \J_2)^{1/2}}$ and differentiating again with $\partial_u$ yields:
\begin{equation}
  u
 = \frac{T \sqrt{2\widetilde D}}{\big[G_c^{-1}+[\sigma](u)\big]^2} \frac 12 {\J_2}^{-1/2}  \, \J_4\;.
\end{equation}
One finally obtains that $\sigma(t,u)$ can be determined by solving the coupled self-consistent equations
\begin{equation}
  \sigma(u)=2\beta\widetilde D \J_0
\quad \text{and}\quad
  u=2^{-\frac 32}\widetilde D^{-\frac 12}T \frac {\J_4 }{{\J_2 }^{\frac 32}}
  \label{eq:effective_equations_toyGVM}
\end{equation}
and the definition \eqref{eq:def-Jk-lambda} which hides the $t$-dependence,
\emph{e.g.} by isolating $\widetilde G-G(u)$ as a function of $u$ from the second equation, and then inferring $\sigma(u)$ from the first.
These equations are valid on segments of ${u \in \left[ 0,1 \right]}$ where $\sigma(u)$ is not constant.

\subsection{Solution for Cauchy wings at large times}
\label{sec:solut-cauchy-wings}

Up to now, the results are valid for a generic toy model correlator $R_{\xitilde}^\text{toy}(t,\lambda)$.
In Ref.~\cite{agoritsas2010} we have worked out the case of a rounded correlator ${R_{\xitilde}^\text{toy}(t,\lambda)=e^{-\lambda^2\xitilde^2}}$ .
In order to add saturation `wings' to this correlator, we consider a Cauchy Ansatz (see paragraph~\ref{ssec:toymodelwings} and Fig.~\ref{fig:wingedcorrelator}):
\begin{equation}
  R_{\xitilde}^\text{toy}(t,\lambda)=\frac{\lambda^2}{\lambda^2+\ell_t^{-2}} e^{-\lambda^2\xitilde^2}
\end{equation}
where $\ell_t$ corresponds to the transverse lengthscale at which the
wings start to develop. From the infinite-time
result~\eqref{eq:Cxicorr}, we expect $\ell_t$ to diverge as $t$ goes
to infinity, which allows us to study the large scale regime by focusing
on a small $\ell_t^{-1}$ expansion.
%
%
However, bluntly expanding
$R_{\xitilde}^\text{toy}(t,\lambda)$ in powers of $\ell_t^{-1}\ll 1$
in~\eqref{eq:extr_eq_sigma_toymodel} yields divergent integrals; an
alternative solution is rather to solve the coupled
equations~\eqref{eq:effective_equations_toyGVM} perturbatively in
$\ell_t^{-1}$. Expanding $\J_2$ and $\J_4$ in powers of $\ell_t^{-1}$
one finds that the following equation in $X$:
\begin{equation}
  u= 2^{-\frac 12}\widetilde D^{\frac 12} \left[3\Big(\frac{\pi}{X}\Big)^{\frac 14}+7\ell_t^{-2}\pi^{\frac 14}X^{\frac 34}\right]
\end{equation}
admits ${X=\xitilde^2+T\left[ \widetilde G(t)-G(t,u) \right]}$ as a solution to the first order in $\ell_t^{-1}$.
%
The physical solution must go in the limit $\ell_t^{-1}\to 0$ to the
solution obtained in Ref.~\cite{agoritsas2010} at $\ell_t^{-1}=0$,
which involves two RSB cutoffs ${u_*(t) \leq u_c (\xitilde)}$ such
that the solution $\sigma(t,u)$ is non-constant only for ${u_*(t) \leq
  u \leq u_c(\xitilde)}$ and times below the Larkin time $t_c$
defined by ${u_*(t_c) = u_c(\xitilde)}$.
We have actually:
\begin{equation}
\xitilde^2+T \left[ \widetilde G(t)-G(t,u) \right] =\frac{3^4\pi T^4}{2^8 \widetilde D^2u^4} + \frac{3^7 7 \pi^2 T^8}{2^{10}\widetilde D^4 u^8}\ell_t^{-2}+O(\ell_t^{-4})
\end{equation}
Inserting this result into the first relation in~\eqref{eq:effective_equations_toyGVM}, one finds that
\begin{equation}
 \sigma(t,u)
  = \frac{2}{\pi} \frac{\widetilde{D}^2}{T} \left( \frac23 \frac{u}{T} \right)^2 + O(\ell_t^{-2})
\end{equation}
whenever $\sigma(t,u)$ is not constant.
To minimal order in $\ell_t^{-1}$, $\sigma(t,u)$ is translated by the constant $-\frac{\widetilde D }{T}\ell_t^{-1}$.
We also see from its definition \eqref{eq:replica-02} that $[\sigma](t,u)$ is not modified at this order:
\begin{equation}
\begin{split}
  [\sigma](t,u)
  &= [\sigma](t,u)\big|_{\ell_t^{-1}=0}+O(\ell_t^{-2}) \\
  &= \frac{2}{\pi}\widetilde{D}^2 \left( \frac23 \frac{u}{T} \right)^3 + O(\ell_t^{-2})\:.
\end{split}
\end{equation}

Besides, the definition \eqref{eq:replica-02} imposes the condition $[\sigma](t,0)=0$ and thus the existence of $u_*(t)$ for which
\begin{equation}
  [\sigma](t,u\leq u_*(t))=0
\end{equation}
Both $\sigma(t,u)$ and $[\sigma](t,u)$ are constant in $u$ below ${u_*(t)}$ and can be strictly monotonous only above ${u_*(t)}$;
this implies by continuity of $\sigma(t,u)$ that
\begin{equation}
\begin{split}
 \sigma(t,0)
  & =\sigma(t,u\leq u_*(t)) \\
  &=\sigma(t,0)|_{\ell_t^{-1}=0}-\frac{\widetilde D }{T}\ell_t^{-1} + O(\ell_t^{-2})
 \end{split}
\end{equation}
To determine whether the second threshold $u_c(\xitilde)$ has a correction which depends on $\ell_t^{-1}$ at this order, one has to check that the original equation~\eqref{eq:extr_eq_sigma_toymodel} is satisfied.
Using the replica inversion formula \eqref{eq:replica-04},
one checks that $u_c$ being fixed this quantity is not modified at order $O(\ell_t^{-1})$:
\begin{equation}
\widetilde{G} (t) - G(t,u)=\big[ \widetilde{G} (t) - G(t,u)\big]\Big|_{\ell_t^{-1}=0} + O(\ell_t^{-2})
\end{equation}
and replacing its value in the original self-consistent equation for ${\sigma(t,u)}$ \eqref{eq:extr_eq_sigma_toymodel} we see that the equation on $u_c(\xitilde)$ is not modified at the same order.

We can hence compute the modification to the roughness ${B(t)= T \lim_{n \to 0} \widetilde G (t)}$ as given in \eqref{eq:def-roughness-replica}
through the formula \eqref{eq:replica-03}.
Observing that
\begin{align}
  \widetilde G (t) &= 
    \widetilde G(t) \big|_{\ell_t^{-1}=0}
    +\frac{\sigma(t,0)-\sigma(t,0)|_{\ell_t^{-1}=0}}{\left[ G_c^{-1}(t)\right]^2} + O(\ell_t^{-2})
\\ &=
   \widetilde G(t) \big|_{\ell_t^{-1}=0} - \frac{t^2}{c^2} \frac{\widetilde D}{T}\ell_t^{-1} + O(\ell_t^{-2})
\end{align}
we finally obtain 
\begin{align}
  B(t) &=  B(t)\big|_{\ell_t^{-1}=0} - \frac{t^2\widetilde D}{c^2} \ell_t^{-1}  + O(\ell_t^{-2})
\label{eq:Bt_perturbation_GVMtoy_Dtilde} 
\end{align}
At asymptotically large times, we know\cite{agoritsas2010} that
\begin{equation}
  B(t)\big|_{\ell_t^{-1}=0} 
  = \frac 32 \Big(\frac{2\widetilde D^2}{\pi c^4}\Big)^{ \frac 13} t^{\frac 43}-\xitilde^2
\end{equation}
so in the random-manifold roughness regime ${B_{\text{RM}}(t) \sim \frac 32 \Big(\frac{2\widetilde D^2}{\pi c^4}\Big)^{ \frac 13} t^{\frac 43}}$.
Thus for the  lengthscale $\ell_t$ not to modify the scaling of this
asymptotic behavior via the correction ${- \frac{t^2\widetilde D}{c^2}
  \ell_t^{-1} }$, we must have consistently ${\ell_t \gtrsim
  \sqrt{B_{\text{RM}}(t)}}$.

%

%

\section{From the discrete to the continuous DP}
\label{app:discretetocontinumDP}

In this appendix, we determine formally the continuum limit of a
discrete directed polymer model, whose parameters are denoted for
convenience in Gothic script (the inverse temperature $\betag$, the
disorder strength $\Dg$, and when needed the elastic constant
$\cg$). See also Ref.~\cite{bustingorry2010} for a detailed analysis of the
scalings and Ref.~\cite{alberts_continuum_2012} for a mathematical
approach.

\subsection{Simplest case: jump to the two nearest neighbors ($n=1$)}

The partition function $Z_{t,y}$ of the discrete SOS model, where the polymer
can jump one step either to its right or to its left at each time step, obeys the 
recursion relation
\begin{equation}
  Z_{t,y}=
  e^{-\betag V_{t,y} }
  \Big[Z_{t-1,y-1}+Z_{t-1,y+1}\Big]
\end{equation}
It is described by an inverse temperature $\betag$ and a disorder
strength $\Dg$ (hidden in $V$ through $\overline{V_{t,y}V_{t',y'}}=\Dg \delta_{tt'}\delta_{yy'}$).  To explicit the
correspondence between the discrete parameters $\betag$ and $\mathfrak
{D} $ and the continuum parameters $\beta$, $c$ and $D$ (\emph{e.g.} of the evolution
equation~\eqref{eq:eqevolWpolym}), we may explicit the lattice
spacings $\pa$ and $\pb$ in directions $y$ and $t$ respectively:
\begin{equation}
  Z_{t,y}=
  e^{-\betag (\pa\pb\Dg)^{\frac 12} V^1_{t,y} }
  \Big[Z_{t-\pb,y-\pa}+Z_{t-\pb,y+\pa}\Big]
\end{equation}
where the factor $(\pa\pb\Dg)^{\frac 12}$ in front of the disorder
is chosen so that $V^1_{t,y}\equiv (\pa\pb\Dg)^{-\frac 12}V_{t,y}$
becomes a white noise in the continuum limit:
\begin{align}
    \overline{V^1_{t,y}V^1_{t',y'}} &= \frac{1}{\pa\pb\Dg} \overline{V_{t,y} V_{t',y'}} 
    = \frac 1\pb\delta_{t,t'}\:\frac 1\pa \delta_{y,y'} \\&\xrightarrow{\pa,\pb \to 0} \delta(t-t')\delta(y-y')
\end{align}

Introducing for normalization purposes ${W_{t,y}=2^{-t}Z_{t,y}}$\,, one expands its corresponding equation of evolution as follows:
\begin{align}
  W_{t,y}&= \frac 12  e^{-\betag (\pa\pb\Dg)^{\frac 12} V^1_{t,y}} \Big[W_{t-\pb,y-\pa}+W_{t-\pb,y+\pa}\Big]
\\
  &\simeq \frac 12  \big[1-\betag (\pa\pb\Dg)^{\frac 12} V^1_{t,y}\big] \Big[2W_{t-\pb,y}+\pa^2\partial_y^2W_{t,y}\Big]
\end{align}
so that, with $W_{t,y}-W_{t-\pb,y}\simeq \pb\partial_tW_{t,y}$
\begin{align}
 \partial_tW_{t,y}&= \frac 12 \frac{\pa^2}\pb \partial_y^2W_{t,y} -\betag (\pa\pb^{-1}\Dg)^{\frac 12} V^1_{t,y} W_{t,y}
\end{align}
which corresponds for instance to the continuum  Feynman-Kac  evolution~\eqref{eq:eqevolWpolym} with the parameters
\begin{equation}
  \beta=\betag \qquad c=\frac{\pb}{\pa^2}\frac1{\betag} \qquad D=\pa\pb^{-1}\Dg
  \label{eq:cbetaDmicmacn1}
\end{equation}
This correspondence between discrete and continuum parameters is used
in section~\ref{sec:numericalresults_discrete} when discussing
numerical results on the discrete DP.

\subsection{Generic case: jump to the $2n$ nearest neighbors }

For the sake of completeness, we now consider a generalized SOS model
in the spirit of Ref.~\cite{bustingorry2010}, where the polymer endpoint $y$
can jump to either of its $2n$ neighbors $y-2n+1,y-2n+3,\ldots,y+2n-1$ with an elastic weight 
depending on the distance $j$ as $e^{-\betag j^2}$.
The partition function now obeys 
\begin{equation}
  Z_{t,y}=
  e^{-\betag V_{t,y}}
  \sum_{j=1}^{n}e^{-\betag j^2}\Big[Z_{t-1,y-(2j-1)}+Z_{t-1,y+(2j-1)}\Big]
\end{equation}
We explicit as previously the lattice spacings $\pa$ and
$\pb$, and we also introduce a microscopic elastic constant $\cg$
\begin{align}
  Z_{t,y}&=
  e^{-\betag(\pa\pb\Dg)^{\frac 12} V^1_{t,y}} \times
 \\ &\qquad
  \sum_{j=1}^{n}e^{-\frac 12 \cg \pa^2 \betag j^2}\Big[Z_{t-\pb, y-(2j-1)\pa}+Z_{t-\pb,y+(2j-1)\pa}\Big]  
  \nonumber
\end{align}

Introducing, for normalization purposes,
$W_{t,y}=(2\Omega)^{-t}Z_{t,y}$ one expands its corresponding equation
of evolution as follows:
\begin{align}
  W_{t,y}&\simeq \frac 1{2\Omega}  \big[1-\betag(\pa\pb\Dg)^{\frac 12} V^1_{t,y}\big] \times 
\label{eq:eqevolWabn}
  \\ &\qquad
  \sum_{j=1}^{n}e^{-\frac 12 \cg \pa^2\betag j^2}\Big[2W_{t-\pb,y}+(2j-1)^2 \pa^2\partial_y^2W_{t-\pb,y}\Big]
  \nonumber
\end{align}
To ensure that the dominant order of the right hand side is
$W_{t-\pb,y}$, which allows to recognize a time difference, one sets
\begin{equation}
   \Omega=\sum_{j=1}^{n}e^{-\frac 12 \cg \pa^2\betag j^2}
\end{equation}
Then, defining the effective elastic constant $\kappa_n(\cg,\betag)$ as in Ref.~\cite{bustingorry2010}
\begin{equation}
 \kappa_n=\frac{\sum_{j=1}^{n}e^{-\frac 12 \cg \pa^2\betag j^2}}{\betag\sum_{j=1}^{n} (2j-1)^2e^{-\frac 12 \cg \pa^2\betag j^2}}  
\end{equation}
one recovers from~\eqref{eq:eqevolWabn} the equation of evolution
\begin{equation}
 \partial_t W_{t,y} = 
 \frac 1{2\betag \kappa_n}\frac{\pa^2} \pb\partial_y^2 W_{t,y}- \betag (\pa\pb^{-1}\Dg)^{\frac 12} W_{t,y} V_{t,y}
\end{equation}
which corresponds to the continuum evolution~\eqref{eq:eqevolWpolym}
upon the identification
\begin{equation}
  \beta=\betag \qquad c= \frac{\pb}{\pa^2}\kappa_n(\cg,\betag) \qquad D=\pa\pb^{-1}\Dg
  \label{eq:cbetaDmicmac}
\end{equation}
Note that the result holds for any $n$; in particular, $\kappa_n=\frac
1\betag$ for $n$=1 (as obtained in the previous paragraph) but
${\kappa_n}$ goes to another ${\cg}$-dependent limit for $n^2\gg T$,
which enables in particular to study the high-temperature limit of the
discrete directed polymer model~\cite{bustingorry2010}.

%


\end{document}